\documentclass[aps,prd,twocolumn,showpacs,superscriptaddress,nofootinbib,10pt]{revtex4-2}

\usepackage{acronym}
\usepackage{amsfonts}
\usepackage{amsmath}
\usepackage{amssymb}
\usepackage{bm}
\usepackage{cases}
\usepackage{color}
\usepackage{comment}
\usepackage[dvipsnames]{xcolor}
\usepackage{enumerate}
\usepackage{epsfig}
\usepackage{etoolbox}
\usepackage{fancyref}
\usepackage{fancyhdr}
\usepackage[T1]{fontenc} 
\usepackage{graphicx}
\usepackage{hyperref}
\usepackage{indentfirst}
\usepackage{latexsym}
\usepackage{mathrsfs}
\usepackage{mciteplus}
\usepackage{multirow}
\usepackage{rotating}
\usepackage{scrextend}
\usepackage{soul}
\usepackage{subfigure}
\usepackage{verbatim}
\usepackage{amsthm} 

\newcommand{\jeven}{\mbox{\rm\j}}

\begin{document}

\title[Black Hole Greybody Factors from Korteweg-de Vries Integrals: Theory]{Black Hole Greybody Factors from Korteweg-de Vries Integrals: Theory}

\author{Michele Lenzi}
\affiliation{Institut de Ci\`encies de l'Espai (ICE, CSIC), Campus UAB, Carrer de Can Magrans s/n, 08193 Cerdanyola del Vall\`es, Spain}
\affiliation{Institut d'Estudis Espacials de Catalunya (IEEC), Edifici Nexus, Carrer del Gran Capit\`a 2-4, despatx 201, 08034 Barcelona, Spain}

\author{Carlos F. Sopuerta}
\affiliation{Institut de Ci\`encies de l'Espai (ICE, CSIC), Campus UAB, Carrer de Can Magrans s/n, 08193 Cerdanyola del Vall\`es, Spain}
\affiliation{Institut d'Estudis Espacials de Catalunya (IEEC), Edifici Nexus, Carrer del Gran Capit\`a 2-4, despatx 201, 08034 Barcelona, Spain}

\begin{abstract}
The dynamics of perturbed non-rotating black holes (BHs) can be described in terms of master equations of the wave type with a potential. 
In the frequency domain, the master equations become time-independent Schr\"odinger equations with no discrete spectrum. 
It has been recently shown that these wave equations possess an infinite number of symmetries that correspond to the flow of the infinite hierarchy of Korteweg-de Vries (KdV) equations. 
As a consequence, the infinite set of associated conserved quantities, the KdV integrals, are the same for all the different master equations that we can consider. 
In this paper we show that the BH scattering reflection and transmission coefficients characterizing the continuous spectrum can be fully determined via a moment problem, in such a way that the KdV integrals provide the momenta of a distribution function depending only on the reflection coefficient. We also discuss the existence and uniqueness of solutions, strategies to solve the moment problem, and finally show the case of the P\"oschl-Teller potential where all the steps can be carried out analytically.
\end{abstract}

\maketitle

\section{Introduction}

General Relativity predicts the existence of Black Holes (BHs), regions of spacetime from where nothing can escape, not even light~\cite{Hawking:1973uf,Wald:1984cw,Novikov:1989sz,Chandrasekhar:1992bo}. After many years of study of these intriguing objects from many different points of view, some spectacular results have been discovered. Of particular relevance is the no-hair {\em conjecture}, which tells us that isolated BHs are characterized just by three numbers: Mass, spin, and electric charge (not relevant from the astrophysical point of view). An astonishing fact about BHs is that they can come in any size, the only impediment against their existence is to have a viable formation mechanism. BHs are central objects in several areas of Physics and Astronomy.  They also play a role in the development of fundamental physics theories that include gravity and go beyond the standard model of particle physics. Of special relevance are theories that involve higher dimensions or implement the holographic principle.  From the observational point of view, accumulating evidence of BHs (or of objects that look like BHs) come from various sources, in particular X-rays~(see, e.g.~\cite{2016A&A...587A..61C}) and gravitational waves~\cite{LIGOScientific:2018mvr,LIGOScientific:2021usb,LIGOScientific:2021djp}. In the case of gravitational waves, we expect to learn more about BHs with future third-generation detectors~\cite{Sathyaprakash:2012jk,Evans:2021gyd} and space-based detectors like LISA~\cite{LISA:2017pwj,Amaro-Seoane:2022rxf,LISACosmologyWorkingGroup:2022jok,LISA:2022kgy}.

Among the possible processes involving BHs, there are two of great relevance: Scattering processes and quasinormal mode (QNM) oscillations. We devote this paper to the first kind of processes. Scattering theory for BHs~\cite{Futterman:1988ni,Andersson:2000tf} has seen a number of resurgences during time in line with different developments in astrophysics and fundamental physics. This is not surprising since scattering methods are in general a powerful tool to investigate the nature of the scatterer by collecting asymptotic data. Therefore, various authors, during decades (see, e.g.~\cite{Vishveshwara:1970zz,Starobinskil:1974nkd,Sanchez:1976fcl,Sanchez:1976xm,Sanchez:1977vz,Futterman:1988ni,Andersson:2000tf,Castro:2013lba,Folacci:2019vtt}), have explored this field looking for details on the nature of the gravitational interaction and the implications for gravitational quantum theory.

In the case that the scattered particles or waves can be considered as small perturbations of the BH geometry, the best tool to study scattering processes on a non-trivial curved background is spacetime perturbation theory. In this case, BH perturbation theory (BHPT). As it was found in pioneering works in Refs.~\cite{Regge:1957td,Cunningham:1978cp,Cunningham:1979px,Cunningham:1980cp,Zerilli:1970la,Zerilli:1970se,Moncrief:1974vm} (see also Ref.~\cite{Chandrasekhar:1992bo}), and generalized to d-dimensional BHs in Refs.~\cite{Ishibashi:2003ap,Kodama:2003kk}, the physical content of the perturbations around a BH background can be described by master wave-like equations with a potential (and eventually source terms in the case that these perturbations are generated by matter fields). When considered in the frequency domain, such master equations are actually time-independent Schr\"odinger equations for each frequency. Therefore, BHPT together with scattering theory permits to describe a quite large variety of physical phenomena~\cite{Futterman:1988ni}.

Greybody factors are key physical quantities in this context. They are nothing but the transmission (and reflection) probabilities associated with the potential barrier and are therefore highly informative about the nature of the object (a BH in this case, but other more exotic compact object can also be considered) and its response to the perturbations. By construction, greybody factors represent the amount of particles or waves that are transmitted through the BH barrier. This is particularly important, for example, when studying Hawking radiation~\cite{Hawking:1975vcx}, as greybody factors encode the deviations from the black-body spectrum caused by the interaction with the background potential. Furthermore, the poles of the analytic extension of the transmission coefficient represent the discrete complex energies associated to resonances, known as the BH quasi-normal modes (QNMs), that is, the characteristic damped oscillation modes of the BH~\cite{Kokkotas:1999bd,Nollert:1999re,Berti:2009kk,Konoplya:2011qq}. These modes are of crucial importance in the recent era of Gravitational Wave Astronomy, as they can be used as fingerprints of new physics~\cite{Barack:2018yly,Perkins:2020tra,LISA:2022kgy}.

There is substantial literature about calculations of greybody factors. The first calculations date back to the mid 70s where already low-frequency~\cite{Page:1976df,Unruh:1976fm,Starobinsky:1973aij,Starobinskil:1974nkd} and high-frequency~\cite{Sanchez:1976fcl,Sanchez:1976xm,Sanchez:1977vz} limits were presented using matching techniques at the turning points. A powerful technique is the Wentzel–Kramers–Brillouin (WKB) approximation~\cite{Bender:1978bo}, first applied in this context in Refs.~\cite{Schutz:1985km,Iyer:1986np,Iyer:1986nq} and then extended to higher orders in the approximation in Refs.~\cite{Konoplya:2003ii,Matyjasek:2017psv,Konoplya:2019hlu}. Other approaches to the calculation of greybody factors include the bounding of the Bogoliubov coefficients~\cite{Visser:1998ke,Boonserm:2008zg}, monodromy methods~\cite{Neitzke:2003mz,Castro:2013lba,Harmark:2007jy}, etc.  For the case of higher-dimensional BHs see~\cite{Harmark:2007jy}.

In this paper, we use recent results on the structure of BHPT~\cite{Lenzi:2021wpc,Lenzi:2021njy} to derive a new method for the computation of greybody factors. In~\cite{Lenzi:2021wpc}, the full space of master functions and equations for the perturbations of Schwarzschild BHs was constructed. Two branches of master equations were distinguished: (i) The {\em standard} branch, which contains master equations with the known potentials, namely the Regge-Wheeler potential~\cite{Regge:1957td} for odd-parity perturbations and the Zerilli potential~\cite{Zerilli:1970la} for even-parity perturbations.(ii) The {\em Darboux} branch, where there is an infinite set of master equations with new potentials.  In~\cite{Lenzi:2021njy}, it was found that all these master equations can be linked by means of Darboux transformations (see also~\cite{Glampedakis:2017rar}), establishing in this way their physical equivalence given that the Darboux transformation preserves the spectrum, that is, it is isospectral. This means that all these master equations lead to the same reflection and transmission coefficients, thus extending the result of Refs.~\cite{1980RSPSA.369..425C,Chandrasekhar:1992bo}. Therefore, Darboux transformations emerge as a symmetry of the master equations describing perturbations of the Schwarzschild geometry, named as {\em Darboux covariance}~\cite{Lenzi:2021njy}.

Moreover, the structure of the space of master functions gets enriched by the introduction~\cite{Lenzi:2021njy} of inverse scattering techniques~\cite{Faddeev:1959yc, Faddeev:1976xar, Deift:1979dt, Novikov:1984id, Ablowitz:1981jq}. These techniques are used to solve non-linear evolution problems~\cite{Miura:1968JMP.....9.1202M,Gardner:1967wc}, like for instance the Korteweg-de Vries (KdV) equation, which describes a wide range of physical phenomena~\cite{doi:10.1137/1018076}. The key idea behind the inverse scattering method is to reconstruct the time evolution of a potential (the unknown satisfying the non-linear evolution equation) from the evolution of its scattering data, which is obtained from an associated linear and time-independent problem. In the case of the KdV equation, the linear problem is the time-independent Schr\"odinger equation. Another remarkable result is the appearance of an infinite series of conservation laws for the KdV equation with the corresponding set of conserved quantities~\cite{Miura:1968JMP.....9.1204M, Zakharov:1971faa, Lax:1968fm}, the so-called KdV integrals. We can apply these techniques to our master equations in the frequency domain and generate the associated set of KdV integrals~\cite{Lenzi:2021njy}. One can see that the KdV equation constitutes an isospectral deformation of the master equations so that the transmission coefficient and the quasi-normal modes are preserved by the KdV deformation.  More importantly, it was shown in~\cite{Lenzi:2021njy} that the KdV integrals are invariant under Darboux transformations. 

~

\noindent{\em Executive Summary}. In this paper we show that the KdV integrals associated with the BH effective potential fully determine the transmission probability, or greybody factors, which contains all the relevant physical information about BH scattering processes. The way in which the KdV integrals determine the transmission probability is through a {\em moment problem}, that is, the KdV integrals turn out to be the moments (up to a trivial multiplicative factor) of a distribution function associated with the transmission probability. We also show that this moment problem is determinate, so that a solution exists and it is unique. This means that it is possible to invert the moment problem to find the greybody factors. We discuss possible avenues to carry out the inversion and illustrate it with an example that can be analytically solved: The particular case of a P\"oschl-Teller potential.

~

\noindent{\em Structure of the Paper}. In Section~\ref{bhpt-summary} we introduce BHPT for the case of a Schwarzschild BH and introduce the main elements that are necessary for the developments of this work. In Section~\ref{master-landscape} we review the landscape of possible master functions and equations describing the BH perturbations. In Section~\ref{section-darboux-covariance} we show how Darboux transformations work and their consequences. In Section~\ref{Sec:kdv-symmetries}, we introduce the concept of KdV deformation of the master equations, their structure and how they generate an infinite sequence of conservation laws and conserved quantities, the KdV integrals. We also show their invariance under Darboux transformations. In Section~\ref{bh-scattering}, using all the previous tools, we study BH scattering and arrive to the main result of this work: The formulation of a moment problem for the transmission coefficient, where the moments are essentially the KdV integrals associated with the BH potential(s). In Section~\ref{Sec:moment-problem}, we discuss the inversion of the moment problem and illustrate it in the case of an analytically solvable case. We finish the paper in Section~\ref{Conclusions-and-Discussion} with a discussion of the next steps in this type of study and of future perspectives and potential applications. The paper contains five appendices: Appendix~\ref{gauge-invariant-mp-harmonics} on gauge invariant metric perturbations; Appendix~\ref{Hamiltonian-Formalism} about the Hamiltonian formalism for infinite–dimensional system; Appendix~\ref{App:1d-scattering} on basic elements of one-dimensional scattering processes; Appendix~\ref{App:PT} about some results on the P\"oschl-Teller potential; and Appendix~\ref{App:KdV} shows the first KdV integrals for the potentials associated with a Schwarzschild BH.

~

Throughout this paper, otherwise stated, we use geometric units in which $G = c = 1\,$. For convenience, we use multiple notations for partial derivatives of a given function $f(x)$: $\partial f/\partial x$, $\partial_x f$, $f_{,x}$.

\section{Summary of BHPT in the non-rotating case} \label{bhpt-summary}

The study of perturbations of BHs started with the seminal paper of Regge-Wheeler~\cite{Regge:1957td}, which already laid many of the elements of the theory that are used now, at least in the non-rotating case.  Let us now summarize these ingredients and the last developments that are the basis of the work presented in this paper. For more details see~\cite{Nagar:2005ea,Martel:2005ir,Lenzi:2021wpc}.

\subsection{Formulation of the first-order perturbative equations}
In relativistic perturbation theory we deal with spacetimes that describe phenomena that, in a sense, can be considered as {\em small} deviations of an idealized physical situation/system. This idealized situation is encoded in what we call the {\em background} spacetime, which typically is a spacetime with a certain degree of symmetry. In our case, the background is the Schwarzschild spacetime~\cite{Schwarzschild:1916uq}, i.e. an isolated non-rotating BH, which is static and spherically symmetric.  The metric of the background spacetime has then the following simple form\footnote{We use a hat to denote quantities associated with the background spacetime.}:
\begin{equation}
d\hat{s}{}^2=\widehat{g}^{}_{\mu\nu}dx^{\mu}dx^{\nu}=-f(r)\,dt^2+\frac{dr^2}{f(r)}+r^2d\Omega^2\,,
\label{schwarzschild-metric}
\end{equation}
where
\begin{equation}
f(r) = 1-\frac{r_{s}}{r}\,, 
\end{equation}
where $r_s$ is the location of the event horizon, the Schwarzschild radius: $r_{s} = 2GM/c^{2}=2M\,$.
 
The physical spacetime is a member of a one-parameter family of spacetimes, $(\mathcal{M}_\lambda, g_\lambda)$, which is the way in which perturbation theory is usually formulated (see, e.g.~\cite{Stewart:1974uz,Wald:1984cw}). In this way, the perturbations are defined as the derivative terms of the Taylor series expansion in $\lambda$, evaluated on the background ($\lambda=0$). The parameter $\lambda$ is, in general, a dummy parameter that controls the strength of the perturbations and, for simplicity, we can ignore it. Here, we only consider first-order (order $\lambda^1$) perturbations, $h_{\mu\nu}$ ($|h_{\mu\nu}|\ll|\widehat{g}_{\mu\nu}|$), so that the spacetime metric of the physical spacetime can be written as:
\begin{equation}
g^{}_{\mu\nu} = \widehat{g}^{}_{\mu\nu} + h^{}_{\mu\nu} \,.
\label{fundamental-perturbative-equation}
\end{equation}
We denote the first-order perturbation of any quantity ${\cal Q}$ as: $\delta Q = Q - \widehat{Q}$. In particular, $h_{\mu\nu} = g_{\mu\nu} - \widehat{g}_{\mu\nu} = \delta g_{\mu\nu}$. 
 
We can introduce Eq.~\eqref{fundamental-perturbative-equation} into the full Einstein equations and, keeping only terms linear in the metric perturbations $h_{\mu\nu}$, we obtain the equations for the perturbations, which schematically can be represented as:
\begin{equation}
\delta G[h^{}_{}]^{}_{\mu\nu} = 0 \,, \label{first-order-eqs}
\end{equation}
where $\delta G$ is a linear operator that when applied to the perturbations yields the first-order perturbative equations (in vacuum).  These equations are partial differential equations (PDEs) that are coupled. A natural strategy is to try to solve them by looking for combinations of the metric perturbations (and their derivatives) so that they satisfy equations that are decoupled from the rest of variables. These type of combinations and their corresponding equations are usually known as master variables and equations respectively. Of course, the success of this strategy depends on the characteristics of the background spacetime.

Spacetime perturbations are subject to a gauge freedom that comes from the infinite ways we have to establish a correspondence between the background and physical spacetimes~\cite{Stewart:1974uz,Bruni:1996im}. A gauge transformation of the perturbations relates two of these correspondences. In practice, by pulling back the physical metric and the geometric structure associated with it to the background spacetime, a gauge transformation connects two points of the background spacetime in the form of a coordinate transformation: 
\begin{equation}
x^{\mu}\quad\longrightarrow\quad  x'^{\mu} = {x}^{\mu} + \xi^{\mu}\,,
\label{gauge-transformation}
\end{equation}
in such a way that ${x}^{\mu}$ and $x'^{\mu}$ are the coordinates of two points of the background spacetime that correspond to a single point of the physical spacetime.  Here, the vector field $\xi^{\mu}$ ($|\xi^{\mu}|\ll|\widehat{g}_{\mu\nu}|$), is the generator of the gauge transformation. The change induced in the metric perturbations looks as:  
\begin{equation}
h^{}_{\mu\nu}\quad\longrightarrow\quad {h'}^{}_{\mu\nu} = h^{}_{\mu\nu}  -2\,\xi^{}_{(\mu ;\nu)}\,.
\label{gauge-transformation-metric}
\end{equation}
%

\subsection{Spherically Symmetric Background}

The spherical symmetry of the Schwarzchild metric~\eqref{schwarzschild-metric} tells us that it can be decomposed as the warped product of two bi-dimensional geometries ($\mathcal{M}_0 = M_2 \times_r S^2$):
\begin{equation}
d\hat{s}{}^2 = g^{}_{ab}(x^c)dx^a dx^b + r^2(x^a) \Omega^{}_{AB}(x^C) dx^A dx^B\,,
\label{metric-schwarzschild-2}
\end{equation}
where $g^{}_{ab}$ is a Lorentzian geometry [we can choose the coordinates to be $x^{a}=(t,r)$] and $\Omega_{AB}$ is the metric of the 2-sphere [we can choose the coordinates to be $\Theta^{A}=(\theta,\varphi)$]. In the case of the Schwarzschild metric we can write
\begin{eqnarray}
g^{}_{ab} dx^a dx^b & = & -f(r)\,dt^2+\frac{dr^2}{f(r)} \,, \\
\Omega^{}_{AB} d\Theta^A d\Theta^B & = & d\theta^2 +\sin^2\theta\, d\varphi^2 \,. \label{warped-product-schwarzschild}
\end{eqnarray}
In Eq.~\eqref{metric-schwarzschild-2}, $r(x^a)$ is the area radial coordinate and $r^2$ is the geometry warp factor. The covariant derivative on $S^2$ is denoted by a vertical bar ($\Omega_{AB|C} = 0$) while for $M^2$ we use a colon ($g_{ab:c} = 0$). The volume Levi-Civita tensor in $S^2$ is denoted by $\epsilon_{AB}$ while in $M^{2}$ is denoted by $\varepsilon_{ab}\,$. 

This special geometric structure allows for the separation, e.g. for solutions of the wave equation, of the dependence on the coordinates of $M^2$ from the angular dependence. Then, we can expand the solution in spherical harmonics, or when applied to the Einstein perturbative equations, in tensor spherical harmonics. It is well-known that the different harmonics can be divided according to how they transform under parity transformations [$(\theta,\phi)$ $\rightarrow$ $(\pi-\theta, \phi+\pi)$]. In this way, we can distinguish even-parity harmonics, ${\cal E}^{\ell m}$, which transform as: ${\cal E}^{\ell m} \rightarrow (-1)^{\ell}{\cal E}^{\ell m}$, and odd-parity harmonics, ${\cal O}^{\ell m}$, which transform as ${\cal O}^{\ell m} \rightarrow (-1)^{\ell+1}{\cal O}^{\ell m}$.
The basis of scalar, vector and tensor spherical harmonics that are needed are: (i) scalar harmonics $Y^{\ell m}$, which are eigenfunctions of the Laplace operator on $S^2$:
\begin{eqnarray}
\Omega^{AB}Y^{\ell m}_{|AB} = -\ell( \ell +1)Y^{\ell m}\,.
\end{eqnarray}
(ii) Vector spherical harmonics (for $\ell \geqslant 1$). In the even-parity case they are $Y^{\ell m}_{A} \equiv Y^{\ell m}_{|A}$ and in the odd-parity case: $X^{\ell m}_{A} \equiv -\epsilon^{}_A{}^B Y^{\ell m}_{B}\,$. (iii) Symmetric  tensor spherical harmonics ($2$nd-rank, for $\ell \geqslant 2$). In the even-parity case we have $T_{AB}^{\ell m} \equiv Y^{\ell m}\,\Omega_{AB}$ (trace) and $Y^{\ell m}_{AB} \equiv Y_{|AB}^{\ell m} + (\ell(\ell+1)/2)\,Y^{\ell m}\,\Omega_{AB}$ (traceless), and for the odd-parity case: $X_{AB}^{\ell m} \equiv X^{\ell m}_{(A|B)}$ (traceless). See~\cite{Martel:2005ir,Lenzi:2021wpc} for more details.

\subsection{Master Equations for each Harmonic of the Metric Perturbations}

The metric perturbations can be written as a multipolar expansion using the basis of scalar ($Y^{\ell m}$), vector ($Y^{\ell m}_A$, $X^{\ell m}_A$), and tensor spherical harmonics ($T_{AB}^{\ell m}$, $Y^{\ell m}_{AB}$, $X^{\ell m}_{AB}$).
In this way, and thanks to the spherical symmetry of the background, the different harmonics decouple, and the harmonics with different parity decouple as well. That is, from the Einstein perturbative equations we obtain separate equations for each $(\ell,m)$ and parity mode (see, e.g.~\cite{Gerlach:1979rw, Gerlach:1980tx}). That is, we can write
\begin{eqnarray}
h_{\mu\nu} = \sum_{\ell ,m} h^{\ell m, \rm odd}_{\mu\nu} + h^{\ell m, \rm even}_{\mu\nu} \,,
\label{eq:2.7}
\end{eqnarray}
and, dropping for simplicity the harmonic numbers $(\ell,m)$,
\begin{equation}
\begin{split}
h^{\rm odd}_{ab} & = 0\,, \\
h^{\rm odd}_{aA} & = h^{}_a X^{}_A\,, \\
h^{\rm odd}_{AB} & = h^{}_2 X^{}_{AB}\,,
\end{split}
\quad 
\begin{split}
h^{\rm even}_{ab} & = h_{ab}^{} Y \,,\\    
h^{\rm even}_{aA} & = \jeven^{}_a Y^{}_A\,, \\
h^{\rm even}_{AB} & = r^2\left( K^{} T^{}_{AB} + G^{} Y^{}_{AB}\right)\,.
\end{split}
\end{equation}
Here\footnote{The distinction of scalar, vector and tensor modes in~\cite{Lenzi:2021wpc} contains typos.}, for each $(\ell,m)$, $h_{ab}$ are scalar perturbations; $(h_a,\jeven_a)$ are vector perturbations; and $(h_2,K,G)$ are the tensor ones, and all depend on the coordinates $\{x^a\}$ of $M^2$.  

As a consequence of the perturbative gauge freedom [see Eqs.~\eqref{gauge-transformation} and~\eqref{gauge-transformation-metric}], there is also a gauge freedom for each harmonic component of the perturbations. See appendix~\ref{gauge-invariant-mp-harmonics} for the list of gauge invariant combinations for each harmonic component of the metric perturbations.

Once we have decomposed the Einstein perturbative equations according to each harmonic mode $(\ell,m)$ and parity we can try to manipulate them in order to find combinations of them, known as {\em master equations}, that isolate certain combinations of the metric perturbations, known as {\em master functions}. The form of the master equations is as follows:
\begin{equation}
\left(-\frac{\partial^2}{\partial t^2} + \frac{\partial^2}{\partial x^2} - V^{\rm even/odd}_\ell  \right)\Psi^{\rm even/odd} = 0\,,
\label{master-wave-equation}
\end{equation}
where $x$ is the {\em tortoise} coordinate, which satisfies $dx/dr = 1/f$; $\Psi_{\rm even/odd}(t,r)$ is the even/odd master function; and $V^{\rm even/odd}_\ell(r)$ is the potential. In the case of odd-parity perturbations, two independent master functions were known: The Regge-Wheeler (RW)~\cite{Regge:1957td} and the Cunningham-Price-Moncrief (CPM)~\cite{Cunningham:1978cp,Cunningham:1979px,Cunningham:1980cp} master functions. They can be written in covariant form as~\cite{Martel:2005ir, Lenzi:2021wpc} (see appendix~\ref{gauge-invariant-mp-harmonics}):
\begin{eqnarray}
\Psi^{}_{\text{RW}} & = & \frac{r^{a}}{r}\tilde{h}^{}_{a}\,,
\label{regge-wheeler-master-function}
\\
\Psi^{}_{\rm CPM} & = & \frac{2 r}{(\ell-1)(\ell+2)}\varepsilon^{ab} \left( \tilde{h}^{}_{b:a} -\frac{2}{r}r^{}_{a}\tilde{h}^{}_{b} \right)\,.
\label{cunningham-price-moncrief-master-function}
\end{eqnarray}
In the case of even-parity perturbations we have the master function introduced by Zerilli~\cite{Zerilli:1970la} and later by Moncrief~\cite{Moncrief:1974vm}. In covariant form (see appendix~\ref{gauge-invariant-mp-harmonics}):
\begin{equation}
\Psi^{}_{\rm ZM} = \frac{2 r}{\ell(\ell+1)}\left\{ \tilde{K} + \frac{2}{\lambda}\left(r^{a}r^{b}\tilde{h}^{}_{ab} - r r^{a}\tilde{K}^{}_{:a}\right) \right\}\,,
\label{zerilli-moncrief-master-function}
\end{equation}
where
\begin{eqnarray}
\lambda(r) =  (\ell-1)(\ell+2) + \frac{3r^{}_{s}}{r}\,. 
\label{lambda-def-sch}
\end{eqnarray}
It is clear from their expressions that these master functions are gauge invariant (see appendix~\ref{gauge-invariant-mp-harmonics}). On the other hand, the potential for odd-parity perturbations is the Regge-Wheeler potential
\begin{equation}
V_{\ell}^{\rm RW}(r) = \left(1-\frac{r^{}_s}{r}\right)  \left(\frac{\ell(\ell+1)}{r^{2}} - \frac{3r^{}_{s}}{r^{3}} \right)  \,,
\label{schwarzschild-omega-potential-odd-parity}
\end{equation}
while for even parity we have the Zerilli potential
\begin{widetext}
\begin{equation}
V^{\rm Z}_\ell(r) = \frac{f(r)}{\lambda^{2}(r)}\left[ \frac{(\ell-1)^{2}(\ell+2)^{2}}{r^{2}}\left( \ell(\ell+1) + \frac{3r^{}_{s}}{r} \right)  + \frac{9r^{2}_{s}}{r^{4}}\left( (\ell-1)(\ell+2) + \frac{r^{}_{s}}{r} \right) \right] \,.
\label{schwarzschild-omega-potential-even-parity}
\end{equation}
\end{widetext}
Once the master equations are solved, in certain gauges, for instance in the Regge-Wheeler gauge~\cite{Regge:1957td}, we can reconstruct all the master perturbations from the master functions.

\section{The Full Landscape of Master Functions and Equations}\label{master-landscape}

In~\cite{Lenzi:2021wpc}, the following question was asked\footnote{Actually, this question was considered in a more general context, where a cosmological constant of arbitrary sign was also included.}: What are {\em all} the possible master equations that one can obtain for the vacuum perturbations of a Schwarzschild BH? To answer this question, Ref.~\cite{Lenzi:2021wpc} adopted a general and systematic way of searching for master functions and associated potentials without having to resort to ad hoc combinations of the perturbative field equations that yield decoupled master equations. 

The main outcome of the analysis of~\cite{Lenzi:2021wpc} is a complete picture of the space of master functions and equations, or in other words, of the space of pairs $(V_\ell,\Psi)$ that satisfy a wave equation of the form given in Eq.~\eqref{master-wave-equation}. To that end, only two assumptions on the master functions were imposed: (i) The master functions are linear combinations of the metric perturbations and its first-order derivatives (as in the known master functions). (ii) The coefficients of these linear combinations are time independent, i.e. they only depend on the radial area coordinate $r$ (motivated by the static character of the background). The analysis was done in an arbitrary gauge and, as a by-product, the resulting master functions turned out to be gauge invariant. 

The resulting landscape found in~\cite{Lenzi:2021wpc} reveals the existence of two branches of possible pairs of potentials and master functions (master equations), $(V_\ell,\Psi)$. The first branch contains the known cases and hence it is called the {\em standard} branch. For each parity there is just one potential, the Regge-Wheeler (odd-parity) and Zerilli (even-parity) potentials, i.e.
\begin{eqnarray}
{}^{}_{\rm S}V^{\rm odd/even}_\ell =
\left\{ \begin{array}{lc}
V^{\rm RW}_\ell &~~  \mbox{odd parity}\,, \\[2mm]
V^{\rm Z}_\ell  &~~  \mbox{even parity}\,,
\end{array} \right.
\label{potentials-standard-branch}
\end{eqnarray}
In the standard branch, the most general master function can be written as follows
\begin{eqnarray}
{}^{}_{\rm S}\Psi^{\rm odd/even} =
\left\{ \begin{array}{lc}
\mathcal{C}^{}_1 \Psi^{\rm CPM} + \mathcal{C}^{}_2\Psi^{\rm RW} &  \mbox{odd parity}\,, \\[2mm]
\mathcal{C}^{}_1 \Psi^{\rm ZM} + \mathcal{C}^{}_2\Psi^{\rm NE} &  \mbox{even parity}\,,
\end{array} \right.
\end{eqnarray}
where $\mathcal{C}^{}_1$ and $\mathcal{C}^{}_2$ are arbitrary constants. Then, the most general master function in the odd-parity case is a linear combination of the Regge-Wheeler~\eqref{regge-wheeler-master-function} and the Cunningham-Price-Moncrief~\eqref{cunningham-price-moncrief-master-function} master functions. In the even-parity sector,  the most general master function is a linear combination of the Zerilli-Moncrief~\eqref{zerilli-moncrief-master-function} one and a new master function, $\Psi^{}_{\rm NE}$, found in~\cite{Lenzi:2021wpc}, which reads (see appendix~\ref{gauge-invariant-mp-harmonics}):
\begin{equation}
\Psi^{\rm NE}(t,r) = \frac{1}{\lambda(r)}t^a\left(r \tilde{K}^{}_{:a} - \tilde{h}^{}_{ab}r^b \right) \,.
\label{new-even-parity-master-function}
\end{equation}
The master functions $(\Psi^{\rm CPM},\Psi^{\rm RW})$ and $(\Psi^{\rm ZM},\Psi^{\rm NE})$ are related by a time derivative:
\begin{eqnarray}
t^a \Psi^{\rm CPM}_{\,,a} & = & 2\,\Psi^{\rm RW}\,, \\
t^a \Psi^{\rm ZM}_{\,,a} & = & 2\,\Psi^{\rm NE}\,,
\end{eqnarray}
where the time vector $t^a$ on $M_2$ is defined using its Levi-Civita tensor: 
\begin{equation}
t^a = - \varepsilon^{ab}r_b\,.
\label{m2-time-derivative}
\end{equation}

The second branch is completely new, and is called the {\em Darboux} branch due to the emerging symmetry (see Sec.~\ref{section-darboux-covariance}). This branch contains an infinite class of master functions and potentials pairs. The potential can be any solution of the following nonlinear second-order ordinary differential equation: 
\begin{equation}
\left(\frac{\delta{V}^{}_{,x}}{\delta{V}} \right)^{}_{,x}
+ 2 \left(\frac{V^{\rm RW/Z}_{\ell,x}}{\delta{V}} \right)^{}_{,x} - \delta{V} = 0 \,,
\label{potentials-darboux-branch}
\end{equation}
where $\delta{V} = {}_{\rm D}V^{\rm odd/even}_\ell - V^{\rm RW/Z}_\ell$. The master functions in the Darboux branch depend explicitly on the potential. Given a potential, the master functions read~\cite{Lenzi:2021wpc}
\begin{eqnarray}
{}^{}_{\rm D}\Psi^{\rm odd} & = & 
\mathcal{C}^{}_1 \Psi^{\rm CPM}  + \mathcal{C}^{}_2\left( \Sigma^{\rm odd}\, \Psi^{\rm CPM} + \Phi^{\rm odd} \right)\,,
\label{odd-master-function-darboux}
\\[2mm]
{}^{}_{\rm D}\Psi^{\rm even} & = & 
\mathcal{C}^{}_1 \Psi^{\rm ZM}  + \mathcal{C}^{}_2\left( \Sigma^{\rm even}\, \Psi^{\rm ZM} + \Phi^{\rm  even} \right) \,,
\label{even-master-function-darboux}
\end{eqnarray}
where $\Sigma^{\rm odd/even}(r)$ are given by
\begin{eqnarray}
\Sigma^{\rm odd} & = & -\frac{r^{}_s}{2r^2} + \int^{}_x dx'\; {}^{}_{\rm D}V^{\rm odd}_\ell(x') \,,
\\
\Sigma^{\rm even} & = & \frac{1}{\ell(\ell+1)} \left[ \frac{\lambda - (\ell+2)(\ell-1)}{2\,r}  \right. \nonumber \\ 
 & - & \left. \int^{}_x dx'\; {}^{}_{\rm D} V^{\rm even}_\ell(x') \right]
\,.
\end{eqnarray}
As we can see, they contain integrals of the potential. Here, $\Phi^{\rm odd/even}$ are combinations of metric perturbations and their first derivatives, but they are not by themselves master functions. Only their combinations with $\Psi^{\rm CPM/ZM}$ in Eqs.~\eqref{odd-master-function-darboux} and~\eqref{even-master-function-darboux} are master functions~\cite{Lenzi:2021wpc}.

\section{Darboux Covariance of BH Perturbations} \label{section-darboux-covariance}

\begin{figure}[t]
\centering
\includegraphics[width=0.49\textwidth]{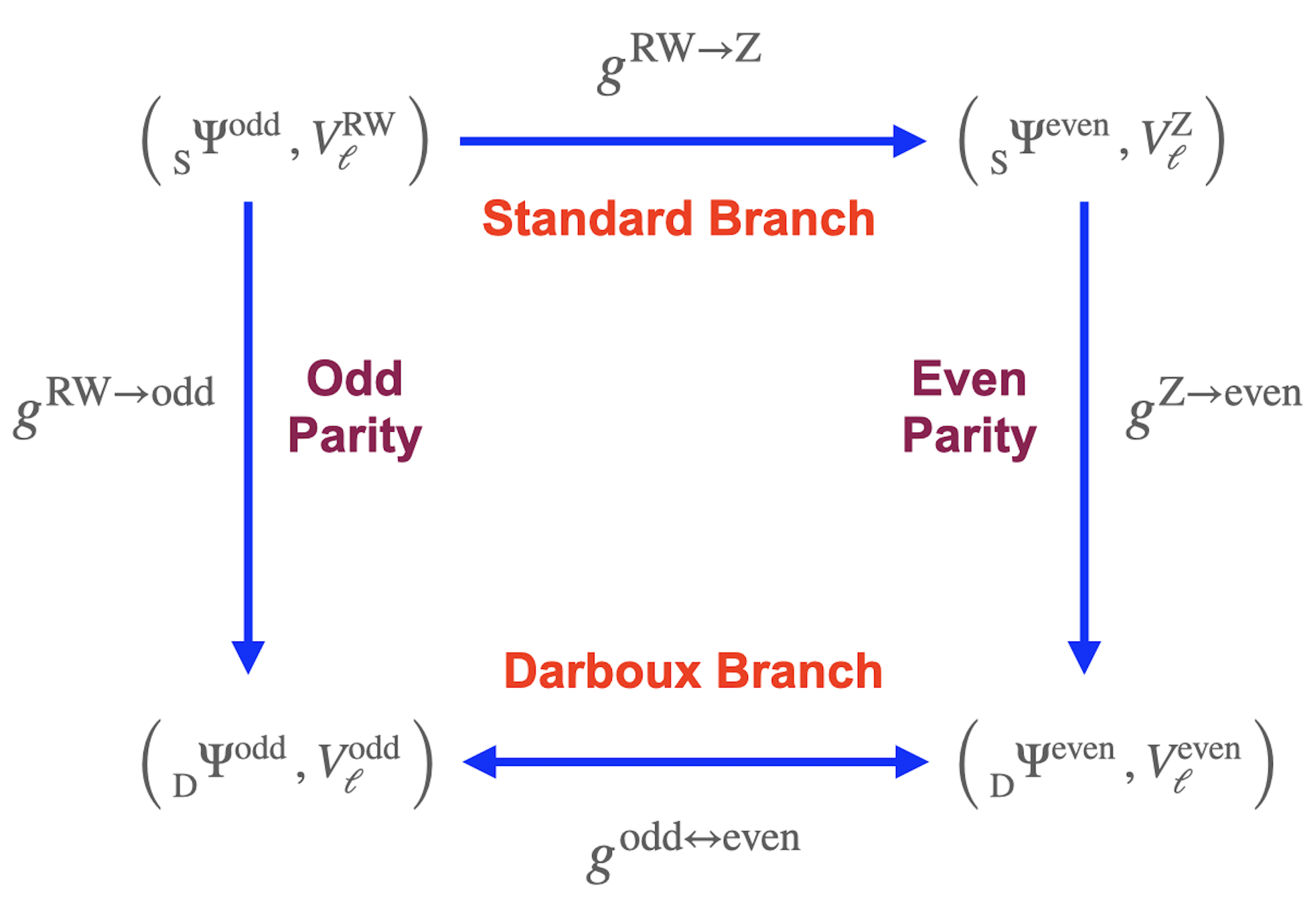}
\caption{Schematic representation of the landscape of master functions and equations for the perturbations of a Schwarzschild BH and how they are linked by DTs.
\label{DT-graphic}}
\end{figure}

The structure and properties of the landscape of master functions and equations was studied in~\cite{Lenzi:2021njy}. The conclusion is that all the pairs $(V,\Psi)$ are related by Darboux transformations (DTs). To understand this, let us consider two such pairs, say  $(v,\Phi)$ and $(V,\Psi)$, that satisfy a wave equation like Eq.~\eqref{master-wave-equation}. We say that they are related by a DT if there is a function $g(x)$, the DT generating function, that maps one pair onto the other as 
\begin{eqnarray}
(v,\Phi)\rightarrow(V,\Psi):\quad
\Psi & = & \Phi^{}_{,x}  + g\,\Phi \,,
\label{Darboux-transformation-wave}
\\
V & = & v + 2\,g^{}_{,x}\,.
\label{Darboux-transformation-potentials}
\end{eqnarray}
The DT generating function $g$ cannot be arbitrary but it has to satisfy the following Riccati equation
\begin{eqnarray}
g^{}_{,x} - g^2 + v = {\cal C} \,,
\label{conditions-on-Darboux-tranformation}
\end{eqnarray}
where ${\cal C}$ is an arbitrary constant. From Eq.~\eqref{Darboux-transformation-potentials} we can isolate $g_{,x}$
\begin{equation}
g^{}_{,x} = \frac{V - v}{2} \,.
\end{equation}
Introducing this into the derivative of the Riccati equation~\eqref{conditions-on-Darboux-tranformation} we obtain
\begin{equation}
g  =  \frac{(V+v)_{,x}}{2 (V- v)}\,,
\end{equation}
Then, the consistency between these expressions for $g(x)$ and $g_{,x}(x)$ yields 
\begin{equation}
\left(\frac{\delta V^{}_{,x}}{\delta V} \right)^{}_{,x}
+ 2 \left(\frac{v^{}_{,x}}{\delta V} \right)^{}_{,x} - \delta V = 0 
\,, 
\label{xdarboux}
\end{equation}
where $\delta V = V-v\,$.  This is precisely equation~\eqref{potentials-darboux-branch}, that is, the equation that any potential in the Darboux branch should satisfy~\cite{Lenzi:2021wpc}, where $v=V^{\rm RW/Z}$. Hence, all master equations in the Darboux branch are connected via a DT to the standard branch, with DT generating functions given by:
\begin{eqnarray}
g^{\rm Z \rightarrow even}_{\rm RW \rightarrow odd} = \frac{1}{2}\int dx\, \left(V^{\rm even}_{\rm odd} - V^{\rm Z}_{\rm RW} \right) \,,
\label{exp-g-odd-and-even}
\end{eqnarray}
while the two parities in the standard branch are connected by a DT with generating function:
\begin{equation}
g^{\rm RW\rightarrow Z} = \frac{1}{2}\int dx\,\left(V^{\rm Z}   - V^{\rm RW} \right) = - g^{\rm Z \rightarrow RW} \,.
\label{exp-g-odd-to-even}
\end{equation}
In conclusion, there is an infinite set of master functions and equations linked by DTs, as illustrated by Figure~\ref{DT-graphic}. This shows the existence of a hidden symmetry in the perturbations of spherically-symmetric (non-rotating) BHs: Darboux covariance~\cite{Lenzi:2021njy}.

The DT preserves the frequency of monocromatic waves, that is, it preserves the spectrum of the frequency-domain operator.  To see this, we can work in the frequency domain, where originally the DT was introduced~\cite{Darboux:89,Crum:10.1093/qmath/6.1.121}. Let us then consider a single-frequency solution, i.e.
\begin{equation}
\Psi(t,r) = e^{ik t}\,\psi(k;x) \,,
\label{time-to-frequency-domain}
\end{equation}
so that the master equation becomes a time-independent Schr\"odinger equation:
\begin{equation}
{\cal L}^{}_V\psi \equiv \psi^{}_{,xx} - V\psi = -k^2\psi \,.
\label{schrodinger}
\end{equation}
One can easily see that DTs map this equation into another Schr\"odinger-type equation with the same eigenvalue $-k^2$,  showing the isospectral character of the DT. 

It is important to mention that the original approach to DTs relies on the existence of a particular solution $\phi_0$, with eigenvalue $-k^2_0$, to build the new master function $\psi$ from the old one, $\phi$ [$\Phi(t,x) = e^{ik t}\,\phi(x)$ like in Eq.~\eqref{time-to-frequency-domain}], via a DT. The new master function can then be written as
\begin{equation}
\psi(x) = \frac{W\left[\phi, \phi^{}_0\right](x)}{\phi^{}_0(x)}
=
\phi(x)_{,x} -\frac{\phi_0(x)_{,x}}{\phi_0(x)} \phi(x)
\,,
\label{darboux-master-frequency}
\end{equation}
where $W$ is the Wronskian, defined as usual 
\begin{eqnarray}
W\left[F^{}_1,F^{}_2\right] =
F^{}_{1,x} F^{}_2 - F^{}_1 F^{}_{2,x} \,.
\label{wronskian-def}
\end{eqnarray}
The DT generating function is given by 
\begin{equation}
g(x) = -(\ln \phi_0)^{}_{,x} \,,
\label{darboux-generating-function}
\end{equation}
and despite $\phi_0$ satisfies the Schr\"odinger equation~\eqref{schrodinger} with eigenvalue $-k^2_0$, both the new and old time-independent master functions, $\psi$ and $\phi$ respectively, satisfy the Schr\"odinger equation~\eqref{schrodinger} with eigenvalue $-k^2$, which shows again the isospectrality of the DT transformation. 

Finally,  It turns out that the Regge-Wheeler equation has an {\em algebraically special} solution~\cite{Chandrasekhar:1984:10.2307/2397739,MaassenvandenBrink:2000iwh} (see~\cite{Glampedakis:2017rar}), namely
\begin{equation}
\phi^{}_0 = \frac{\lambda(r)}{2}\mbox{e}^{-ik^{}_0 x} \,, \quad
k^{}_0 = -i\frac{n^{}_\ell(n^{}_\ell+1)}{3M}\,,
\label{algebraically-special-solution-1}
\end{equation}
where $n^{}_\ell = (\ell+2)(\ell-1)/2$ and $\lambda(r)$ given by  Eq.~\eqref{lambda-def-sch}. Therefore, the DT generating function $g^{{\rm RW}\rightarrow{\rm Z}}$ that goes from the odd- to the even-parity sector of the standard branch (see Figure~\ref{DT-graphic}) can be constructed from this solution. It has the following simple form
\begin{equation}
g^{{\rm RW}\rightarrow{\rm Z}}(x) = i k^{}_0  + \frac{3 r^{}_s f(r)}{r^2\lambda(r)} \,.
\end{equation}
%

\section{Korteweg-de Vries Symmetries}
\label{Sec:kdv-symmetries}

We have described the complete landscape of possible master equations and functions. We have also seen the role of the DT as a way of connecting all of them and establishing the isospectrality of the underlying physical description corresponding to this infinite set. In this sense, the DT plays the role of a covariance or symmetry of this set of possible physical descriptions of (vacuum) BH perturbations.  

There is an independent set of transformations of the master equations that introduces a different kind of symmetry, actually an infinite set of them, and which also have the property of being isospectral. The starting point is the work of Gardner, Greene, Kruskal and Miura~\cite{Gardner:1967wc}, who found a way of solving the Korteweg-de Vries (KdV) equation~\cite{Korteweg:10.1080/14786449508620739}
\begin{equation}
{\rm KdV}[V] \equiv V^{}_{,\tau} - 6 VV^{}_{,x} + V^{}_{,xxx}=0 \,,
\label{kdv-equation}
\end{equation}
by using the inverse scattering method~\cite{Deift:1979dt,Ablowitz:1981jq,Eckhaus:1981tn,Novikov:1984id}.  This discovery constituted a breakthrough in the field that provided a very original way of solving the KdV equation by expressing it in terms of the spectral and scattering theory of the Schr\"odinger operator in Eq.~\eqref{schrodinger}.

Lax~\cite{Lax:1968fm} provided a different point of view. He considered a continuous deformation of the Schr\"odinger operator, $L_V$, which essentially means deforming the potential: 
\begin{equation}
V(x)\rightarrow V(\tau,x)\,,    
\label{KdV-deformation-potential}
\end{equation} 
and introducing the following operator
\begin{eqnarray}
P^{}_V\psi = -4\psi^{}_{,xxx} + 6V\psi^{}_{,x} + 3V^{}_{,x}\psi \,,
\label{Eq2-KdV-like}
\end{eqnarray}
so that 
\begin{equation}
\frac{\partial L^{}_V}{\partial\tau} - \left[P^{}_V,L^{}_V\right] = -{\rm KdV}[V]\cdot {\rm Id} \,.
\label{Lax-pair-equation}
\end{equation}
The operators $(P^{}_V,L^{}_V)$ form a Lax pair, that is a pair of operators whose commutator is the action of an operator corresponding to an integrable non-linear PDE, as the KdV equation, times the unity operator. From Eq.~\eqref{Lax-pair-equation} we deduce that, if $V$ satisfies the KdV equation~\eqref{kdv-equation}, the operator $L_V$ is deformed remaining similar to itself (the different $L_V$ for different $\tau$ are unitary equivalent):
\begin{equation}
L^{}_V(\tau) = U(\tau) \cdot L^{}_V(0) \cdot U^{-1}(\tau)\,,
\end{equation}
where the unitary operator $U(\tau)$ evolves as:
\begin{equation}
\frac{dU(\tau)}{d\tau} = P^{}_V\cdot U(\tau) \,.
\end{equation}
Then, it follows that all the spectral characteristics of the operator $L_V(\tau)$ are preserved by the KdV flow (i.e., the evolution along the {\it time} $\tau$). This establishes an important connection between integrable systems and isospectral problems. 

Actually, if we consider a deformation of our time-independent Schr\"odinger equation~\eqref{schrodinger} consisting of Eq.~\eqref{KdV-deformation-potential} and 
\begin{equation}
\psi(x)\rightarrow \psi(x,\tau)\,, \qquad
k\rightarrow k(\tau) \,,
\end{equation}
and such that $\psi(t,x)$ evolves with $P_V$
\begin{eqnarray}
\psi^{}_{,\tau} = P_V\psi \,,
\label{Eq2-KdV-like-2}
\end{eqnarray}
we have~\cite{Lenzi:2021njy}
\begin{equation}
\left( {\rm KdV}[V]  - (k^{2})^{}_{,\tau} \right)\psi = 0 \,.
\label{KdV-and-eigenvalue-deformation}
\end{equation}
This means that if the potential (the Schr\"odinger equation) is deformed according to the KdV equation we must have
\begin{equation}
\psi\,(k^{2})^{}_{,\tau} = 0\,,
\end{equation}
and therefore $k^2$ is invariant under the KdV flow. This, in particular, applies to the complex quasinormal mode frequencies of the Schwarzschild BH. 

The Lax approach opened the door for a Hamiltonian formulation of the KdV equation and the associated conservation laws.  Gardner~\cite{gardner1971korteweg} was the first to notice that the KdV equation can be written in terms of a Hamiltonian. Zakharov and Faddeev~\cite{Zakharov:1971faa} provided a different point of view of the results of Gardner and collaborators~\cite{Gardner:1967wc,gardner1971korteweg}. They used the Hamiltonian formalism to express the KdV conserved quantities in terms of action-angle variables that appear naturally when looking at the scattering problem associated with the time-independent Schr\"odinger equation (see Section~\ref{bh-scattering}). Actually, we can introduce an infinite hierarchy of KdV evolution equations, associated with the infinite chain of KdV conservation laws, using the Hamiltonian formulation of the KdV equation (see also~\cite{FadeevTakhtajan:1987lda}):
\begin{eqnarray}
\partial^{}_{\tau^{}_k} V = {\cal K}^{}_k[V]\,, \quad k=0,1,2,\ldots
\end{eqnarray}
where we have introduced different evolution parameters $\tau_k$ to illustrate that these equations are independent. The functionals ${\cal K}_k[V]$ are (see Appendix~\ref{Hamiltonian-Formalism}):
\begin{equation}
{\cal K}^{}_k[V] = \left\{ V, {\cal H}^{}_{k+1}[V] \right\}^{}_{\rm GZF} = \frac{\partial}{\partial x} \frac{\delta {\cal H}^{}_{k+1}[V]}{\delta V(x)} \,,
\label{kdv-hierarchy-gzf}
\end{equation}
where the ${\cal H}^{}_k[V]$ constitute an infinite sequence of conserved quantities, which we show below how to construct, that are functionals of $V$:
\begin{equation}
{\cal H}^{}_k[V] = \int^{\infty}_{-\infty} dx\, h^{}_k(V,V^{}_{,x},V^{}_{,xx},\ldots) \,,
\label{kdv-integrals-h}
\end{equation}
and the densities $h^{}_k(V,V^{}_{,x},V^{}_{,xx},\ldots)$ are differential polynomials in the potential $V$. 

In Eq.~\eqref{kdv-hierarchy-gzf}, we are using the Gardner-Zakharov-Faddeev bracket~\cite{gardner1971korteweg,Zakharov:1971faa}. But we can also use the Magri bracket~\cite{Magri:1977gn} (see Appendix~\ref{Hamiltonian-Formalism}) to generate the KdV hierarchy of equations:
\begin{equation}
{\cal K}^{}_k[V] = \left\{ V, {\cal H}^{}_{k}[V] \right\}^{}_{\rm M} = {\cal E}^{}_V \frac{\delta {\cal H}^{}_{k}[V]}{\delta V(x)} \,,
\label{kdv-hierarchy-m}
\end{equation}
where the operator ${\cal E}^{}_V$ is defined in Appendix~\ref{Hamiltonian-Formalism}. Equations~\eqref{kdv-hierarchy-gzf} and~\eqref{kdv-hierarchy-m} show the bi-Hamiltonian character of the hierarchy of KdV evolution equations. Actually, by comparing them we can obtain the Lenard~\cite{Praught:2005SIGMA...1..005P} infinite sequence of recursion relations (see~\cite{gardner1971korteweg,Lax:1968fm}):
\begin{eqnarray}
&& {\cal D}\frac{\delta {\cal H}^{}_{k+1}[V]}{\delta V(x)} = {\cal E}^{}_V \frac{\delta {\cal H}^{}_{k}[V]}{\delta V(x)} \quad \Rightarrow \nonumber \\
&&\frac{\partial}{\partial x} \frac{\delta {\cal H}^{}_{k+1}}{\delta V} = 
\left(-\frac{\partial^3}{\partial x^3} +4V\frac{\partial}{\partial x} + 2 V^{}_{,x}\right) \frac{\delta {\cal H}^{}_{k}}{\delta V} \,.
\end{eqnarray}
All we need to specify is the first density
\begin{equation}
h^{}_0 = \frac{1}{2}V\quad \Rightarrow\quad {\cal H}^{}_0 = \frac{1}{2}\int^{\infty}_{-\infty} dx\, V(x) \,.
\end{equation}
It is easy to show that ${\cal H}^{}_0$ (known as the {\em mean height}) is a conserved quantity by evolving it with respect to $\tau$, using the KdV equation, and using the fact that the integral of total gradients vanishes (by using the decay of $V$ at $x\rightarrow\pm\infty$). Then, from the Lenard recurrence, we get the next conserved quantity:
\begin{equation}
h^{}_1 = \frac{1}{2}V^2\quad \Rightarrow\quad {\cal H}^{}_1 = \frac{1}{2}\int^{\infty}_{-\infty} dx\, V^2(x) \,,
\end{equation}
which is interpreted as the momentum of the KdV wave.  Using the Lenard recurrence again we get~\cite{Whitham:1965gb}
\begin{equation}
h^{}_2 = V^3 + \frac{1}{2}V^2_{,x}~\Rightarrow~{\cal H}^{}_2 = \int^{\infty}_{-\infty}\!\! dx \left(V^3 + \frac{1}{2}V^2_{,x}\right) \,,
\end{equation}
which is interpreted as the energy of the wave. Actually, the KdV equation itself can be written as:
\begin{eqnarray}
\partial^{}_\tau V = {\cal D}\frac{\delta {\cal H}^{}_{2}}{\delta V} 
= {\cal E}^{}_V \frac{\delta {\cal H}^{}_{1}}{\delta V} 
= 6VV^{}_{,x} - V^{}_{,xxx} \,.
\end{eqnarray}
In summary, the KdV dynamics admits two Hamiltonian structures and, under both of them, we have an infinite number of conserved quantities, which we have called ${\cal H}^{}_k[V]$ ($k=0,1,2,\ldots$), and are functionals made out of differential polynomials in the potential $V$. From the different equations, and in particular using the Lenard recursion, we can see that these conserved quantities are in involution, that is, they satisfy:
\begin{equation}
\left\{  {\cal H}^{}_j ,  {\cal H}^{}_k \right\}^{}_{\rm GZF} =
\left\{  {\cal H}^{}_j ,  {\cal H}^{}_k \right\}^{}_{\rm M} = 0\,, \quad j,k = 0,1,2,\ldots
\end{equation}
therefore, we can say that the KdV equation is an integrable system in the classical Liouville sense~\cite{Arnold:1989am}. As a consequence,  Zakharov and Faddeev~\cite{Zakharov:1971faa} were able to carry out the first construction of action-angle variables for the KdV equation (see Sec.~\ref{bh-scattering}). 

To end this section, let us comment on the relation between the KdV flows and the DT. We have seen that the KdV deformation, actually any deformation from the infinite hierarchy of KdV equations, of the time-independent Schr\"odinger equation~\eqref{schrodinger} constitutes an isospectral symmetry of the dynamics of BH perturbations. In addition, there is a strong connection between this set of deformations and the DTs. Following the work of~\cite{Matveev:1991ms} it is possible to show that the system of equations
\begin{equation}
L^{}_V \psi = -k^2\psi\,,\qquad
\partial^{}_\tau\psi = P^{}_V\psi\,, 
\label{schrodinger-deformed}
\end{equation}
is invariant under DTs provided the DT generating function $g$ is KdV-deformed as
\begin{eqnarray}
g^{}_{,\tau} = -g^{}_{,xxx} + 6(V+g^{}_{,x})g^{}_{,x} \,.
\label{kdv-deformation-g}
\end{eqnarray}
%

\section{Black Hole Scattering} \label{bh-scattering}

We have seen the rich structure and multiple properties of the space of possible master functions and equations for the dynamics of BH perturbations. Two different types of isospectral transformation play a key role, namely Darboux covariance and KdV deformations. Under DTs, the potential is transformed via Eq.~\eqref{Darboux-transformation-potentials}, i.e. by a gradient, while under KdV transformations, the potential follows the KdV flow~\eqref{kdv-equation}. 

\begin{figure}[t]
\centering
\includegraphics[width=0.49\textwidth]{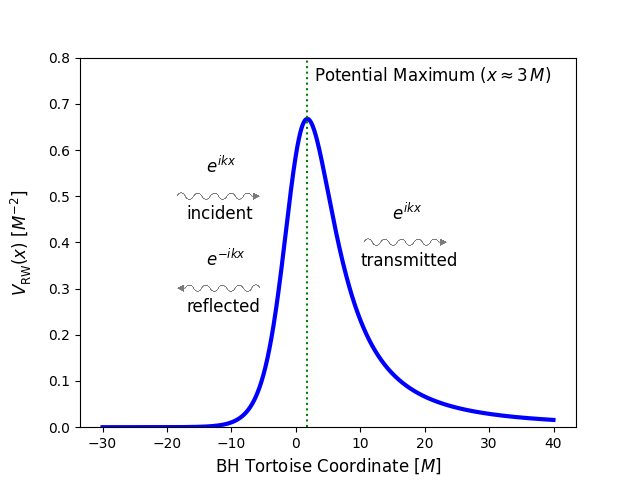}
\caption{Plot of the Regge-Wheeler BH potential~\cite{Regge:1957td} in terms of the tortoise coordinate $x$. The barrier has a maximum at $x\approx 3M=3r_s/2\,$.  This plot is for the particular case $\ell=4\,$. 
\label{BH-scattering-potential} }
\end{figure}

Let us now investigate how we can take advantage of these properties to study BH scattering processes, one of the main applications of BHPT. The typical scattering process is illustrated in Figure~\ref{BH-scattering-potential} where we have a wave packet coming from $-\infty$ (the BH horizon) and, after interacting with the BH potential barrier, there is a reflected part (going towards the horizon) and a transmitted part (going towards spatial infinity). Given that the BH potential barrier is positive, the spectrum of the perturbations contains only the continuum part, that is, there are no discrete bound states. 

We are going to see how to establish a new method to compute the reflection/transmission coefficients that characterize BH scattering. One of the main ingredients is the use of the so-called  trace identities, equations that in our case relate the KdV integrals to the reflection coefficient~\footnote{In the case of the existence of a discrete spectrum, the trace identities also involve the discrete eigenvalues~\cite{Zakharov:1971faa}.} Following the results of~\cite{Lenzi:2021njy} we are going to see that all the possible BH potentials share the same greybody factors as they can be written only in terms of the KdV integrals. In particular, we are going to see that the trace identities can be cast as a symmetric Hamburger moment problem~\cite{schmudgen2017moment}, thus setting the ground for a new method of calculation of greybody factors from KdV integrals.

Let us consider the scattering process illustrated in Fig.~\ref{BH-scattering-potential} and the basics of scattering theory shown in Appendix~\ref{App:1d-scattering} (see also~\cite{Faddeev:1959yc, Faddeev:1976xar, Deift:1979dt, Novikov:1984id, Ablowitz:1981jq} for a detailed treatment). The solution of the time-independent Sch\"odinger equation for this physical situation is given by
\begin{eqnarray}
\psi(x,k,\tau) = \left\{ \begin{array}{lc}
a(k,\tau) e^{i k x} + b(k,\tau)e^{-i k x}  &  \mbox{for~}x\to -\infty\,,  \\
          &                              \\
e^{i k x}  &  \mbox{for~}x \to \infty \,,
\end{array} \right.
\label{plane-wave-ab-deformed}
\end{eqnarray}
where the complex coefficients $a(k,\tau)$ and $b(k,\tau)$ are the so-called Bogoliubov coefficients. They completely determine the transmission/reflection coefficients (and therefore the scattering matrix, see Appendix~\ref{App:1d-scattering}) as follows 
\begin{eqnarray}
t(k,\tau) = \frac{1}{a(k,\tau)} \,, \quad 
r(k,\tau) = \frac{b(k,\tau)}{a(k,\tau)} \,,
\label{reflection-transmission-coefficient-deformed}
\end{eqnarray}
The transmission and reflection probabilities, i.e. the greybody factors, are simply given by the square of the modulus of the corresponding coefficients
\begin{equation}
T(k,\tau) = |t(k,\tau)|^2\,, \quad R(k,\tau)=|r(k,\tau)|^2 \,,
\label{rt-probabilities-deformed}
\end{equation}
which, for real $k$, satisfy Eq.~\eqref{T+R}. Under the KdV flow, the Bogoliubov coefficients evolve as~\cite{Gardner:1967wc}
\begin{eqnarray}
a(k,\tau) = a(k,0) \,,\quad  b(k,\tau) = b(k,0) e^{8 i k^3 \tau} \,,
\label{ab-kdv-flow}
\end{eqnarray}
which means that we have a conservation law
\begin{equation}
a^{}_{,\tau}(k,\tau)=0\,. \label{a-conservation-law}
\end{equation}
The module of $b(k,\tau)$ is also preserved by the KdV flow. Similar evolution laws, can be obtained for the reflection and transmission coefficients. 

When investigating the relations between $r$, $t$ and the KdV integrals, the analytic properties of these functions in the complex plane play a key role~\cite{Newton:1982qc, Taylor:1972pty}. These features depend on the asymptotic behavior of the potential (see Appendix~\ref{App:1d-scattering}), which has to be smooth and integrable over the whole range of $x$. We further require the differential polynomials built out of the potential and its derivatives to be also integrable. It is not difficult to see that all the BH potentials we can obtain via DT transformations~\eqref{Darboux-transformation-potentials} are integrable. Actually, we just have to consider the integrability of the Regge-Wheeler potential together with the asymptotics of all the possible DT generating functions $g$~\cite{Lenzi:2021njy}. Therefore, $a(k, \tau)$ is analytic in the upper half-plane of the complex $k$-plane (see Appendix~\ref{App:1d-scattering} and~\cite{Novikov:1984id}). Then, using the Cauchy theorem, we can write $\ln{a(k, \tau)}$ as
\begin{eqnarray}
\ln{a(k, \tau)} = \frac{1}{2\pi i}
\oint dk'\, \frac{\ln{a(k', \tau)}}{k' - k} \,,
\label{cauchy-a}
\end{eqnarray}
with $\Im(k) > 0$. If we use a typical semicircle contour in the upper complex $k$-plane, the integral over the circular part, thanks to the asymptotic behaviour shown in Eq.~\eqref{asymptotics-a}, goes to zero as we make its radius to be infinite. Therefore, we are left with the integral over the real $k$-line
\begin{eqnarray}
\ln{a(k, \tau)} = \frac{1}{2\pi i}
\int_{-\infty}^{\infty}\!\!\! dk'\, \frac{\ln{a(k', \tau)}}{k' - k} \,.
\label{cauchy-k-line}
\end{eqnarray}
On the other hand, $\overline{a}(k)$ is analytic in the lower half-plane of the complex $k$-plane, and hence we can write the corresponding Cauchy integral using a contour there. 
However, since we are considering $\Im(k) > 0$, the Cauchy transform gives zero as the integrand has no poles in the region of integration, and therefore:
\begin{eqnarray}
\frac{1}{2\pi i} \int_{-\infty}^{\infty}\!\!\! dk'\, \frac{\ln{\overline{a}(k',\tau)}}{k' - k} = 0 \,,
\label{cauchy-a-bar}
\end{eqnarray}
If we now sum Eqs.~\eqref{cauchy-k-line} and~\eqref{cauchy-a-bar}, we obtain
\begin{eqnarray}
\ln{a(k, \tau)} = \frac{1}{2\pi i}\int^{\infty}_{-\infty}\!\!\! dk'\, \frac{\ln T(k',\tau) }{k-k'} \,.
\label{loga-logr}
\end{eqnarray}
Therefore, we can write $a(k, \tau)$, for $\Im(k)\neq 0$, as follows~\cite{Ablowitz:1981jq,Alonso:1982ue}
\begin{eqnarray}
a(k, \tau) & = & \exp\left\{\frac{1}{2\pi i}\int^{\infty}_{-\infty}\!\!\! dk'\, \frac{\ln T(k',\tau) }{k-k'} \right\}\!\!\,, \label{a-full-expression-1} 
\end{eqnarray}
so that, the values in the real line, $\Im(k)=0$, can be obtained as the following limit
\begin{eqnarray}
a(k, \tau) & = & \lim^{}_{\epsilon\rightarrow 0^{+}} a(k+i\epsilon,\tau)\,. \label{a-full-expression-2}
\end{eqnarray}
Then, knowing $a(k, \tau)$ we can find $b(k, \tau)$ from Eq.~\eqref{reflection-transmission-coefficient-deformed}. As a consequence of the asymptotic behavior of $a(k, \tau)$ [see Eq.~\eqref{asymptotics-a}], $\ln a(k, \tau)$ admits, for $|k|\rightarrow\infty$, the following expansion in inverse powers of $k$,
\begin{equation} 
\ln a(k, \tau) = \sum_{n=1}^{\infty} \frac{m^{}_n (\tau)}{k^n} 
\,. 
\label{ak-expansion}
\end{equation}
The trace formulas can then be obtained by evaluating the coefficients $m^{}_n$ of the above expansion in two different ways and then equating them~\cite{Zakharov:1971faa}.
First, we expand Eq.~\eqref{loga-logr} for $\left|k\right|\to \infty$, so that $1/(k - k') \simeq \sum_{n=0}^{\infty} k'^{n}/k^{n+1}$, to obtain
\begin{eqnarray}
\ln a(k, \tau) =
\frac{1}{2\pi i} \sum_{n=0}^{\infty} \frac{1}{k^{2n+1}}
\int_{-\infty}^{\infty}\!\!\! dk'\, k'^{2 n} \ln T(k',\tau) \,.
\end{eqnarray}
This expression has been simplified after integrating away odd functions. As a consequence, only odd coefficients survive so that
\begin{eqnarray}
m^{}_{2n} & = & 0 \,, 
\\ 
m^{}_{2n+1} & = & \frac{1}{2\pi i}
\int_{-\infty}^{\infty}\!\! dk\, k^{2 n} \ln T(k,\tau) \,.
\label{cn-zakharov}
\end{eqnarray}
The second way of calculating the coefficients $m_n$ is based on the Schr\"odinger equation. Let us write the solution in Eq.~\eqref{plane-wave-ab-deformed} as
\begin{equation}
\psi(x,k,\tau) = \exp\left\{i k x - \int_{x}^{+\infty} \!\! dx' \Phi(x',k,\tau) \right\} \,,    
\label{phase-for-psi}
\end{equation}
so that in the half-plane, $\Im(k) > 0$, we have
\begin{eqnarray}
a(k, \tau) =  \lim^{}_{x\rightarrow -\infty} e^{-ik x}\, \psi (x,k,\tau)\,,
\label{a-phi}
\end{eqnarray}
and hence
\begin{eqnarray}
\ln{a(k, \tau) }  = -\int_{-\infty}^{\infty}\!\!\! dx\, \Phi(x,k,\tau) \,.
\label{lna-phi}
\end{eqnarray}
The phase function $\Phi$ satisfies the following complex Riccati equation
\begin{equation}
\Phi^{}_{,x} + 2i k \Phi + \Phi^2 = V \,,
\label{Riccati-equation-complex}
\end{equation}
obtained by substituting Eq.~\eqref{phase-for-psi} into the time-independent Schr\"odinger equation. We can look for solutions of Eq.~\eqref{Riccati-equation-complex} by expanding $\Phi$ in inverse powers of $k$
\begin{equation}
\Phi(x,k, \tau) = \sum_{n=1}^\infty \frac{f^{}_n(x,\tau)}{(2 i k)^n}
\label{expansion-phase}
\,.
\end{equation}
This provides a three-term recurrence relation for the $f_n$ 
\begin{eqnarray}
f^{}_1 & = & V\,, \\
f^{}_n & = & - \frac{d}{dx} f^{}_{n-1} - \sum_{m=1}^{n-1} f^{}_{n-m-1} f^{}_m\,, \; (n = 2, \ldots ) \,.
\label{recurrence-kdv-densities}
\end{eqnarray}
The first coefficients have the form 
\begin{eqnarray}
f^{}_2 & = & - V^{}_{,x}\,, \\
f^{}_3 & = & - V^2 + V^{}_{,xx}\,, \\
f^{}_4 & = & -V^{}_{,xxx} + 4 V V^{}_{,x} \,, \\
f^{}_5 & = & V^{}_{,xxxx} - 6 V V^{}_{,xx} - 5V^{2}_{,x} + 2 V^3\,.
\end{eqnarray}
We see that $f_2$ and $f_4$ are total derivatives. This property holds for all the even coefficients, $f_{2n}$ (see, e.g.~\cite{Lenzi:2021njy}), and therefore, all their integrals vanish. On the other hand, note that each $f^{}_{2n+1}$ is actually proportional, up to total derivative terms, to the densities $h_n$ appearing in the Hamiltonian formalism in Eq.~\eqref{kdv-integrals-h}. The actual relations are
\begin{equation}
2 h^{}_n
=
(-1)^{n} f^{}_{2n +1} + \mbox{Total Derivative Terms} \,,
\end{equation}
where the minus sign is just a consequence of the chosen asymptotics~\eqref{plane-wave-ab-deformed}.
This implies that the functions $f^{}_{2n+1}$ lead to the conserved quantities of the KdV equations~\cite{gardner1971korteweg,Zakharov:1971faa}, i.e. the KdV integrals
\begin{equation}
I^{}_{2n+1}  = \int^{\infty}_{-\infty} \!\! dx\, f^{}_{2n+1}(x)
= 2(-1)^{n} \mathcal{H}^{}_{n} \,.
\label{kdv-integrals}
\end{equation}
Therefore, the odd integrals $I^{}_{2n+1}$, which  are the only non-vanishing ones, actually correspond to the {\em true} first integrals $\mathcal{H}^{}_{n}$~\eqref{kdv-integrals-h} of the KdV equation, i.e. those naturally described by the (bi-)Hamiltonian structure showed in Sec.~\ref{Sec:kdv-symmetries}. Moreover, notice that the odd $I^{}_{n}$ integrals, contrary to the integrals $\mathcal{H}^{}_{n}$, have alternate sign, i.e. $I_1>0, I_3<0, I_5>0\,\ldots$ (see also Appendix~\ref{App:KdV}).

The conservation of the Bogoliubov coefficient $a$ under the KdV flow [see Eq.~\eqref{a-conservation-law}] provides an alternative point of view to the one of Sec.~\ref{Sec:kdv-symmetries} to prove that the integrals $I^{}_n$ are conserved quantities of the KdV equation. In fact, thanks to Eqs.~\eqref{ak-expansion} and~\eqref{expansion-phase} we can check that $d I^{}_n / d\tau =0$. Therefore, in the following we can omit the explicit $\tau$ dependence. However, this is not the only conservation property of the KdV integrals. Indeed, it was shown in~\cite{Lenzi:2021njy} that the infinite set of KdV integrals are also invariant under DTs. Putting the two results together we conclude that the KdV integrals are the same for the whole class of BH potentials. In this sense, we can write
\begin{equation}
I^{}_{2n+1}[V] = I^{}_{2n+1}[V^{}_{\rm RW}] \,,
\end{equation}
where $V$ is any BH potential represented in Figure~\ref{DT-graphic}.

To summarize, from the previous results we conclude that the KdV integrals are really fundamental quantities of BH scattering processes and they naturally reflect the isospectral character of the full space of master equations and functions.

Returning to $\ln a(k)$, we can finally evaluate the coefficients $m^{}_{n}$ in Eq.~\eqref{ak-expansion} by comparing with Eqs.~\eqref{lna-phi},~\eqref{expansion-phase} and~\eqref{kdv-integrals}:
\begin{eqnarray}
m^{}_{2n} & = & 0 \,,  \\ 
m^{}_{2n+1} & = & i\,(-1)^{n}\frac{I_{2n+1}}{2^{2n+1}} \,.
\label{cn-KdV}
\end{eqnarray}
Finally, we match Eqs.~\eqref{cn-zakharov} and~\eqref{cn-KdV} to obtain the following trace formulas
\begin{eqnarray}
(-1)^{n+1}\frac{I_{2n+1}}{2^{2n+1}} =
\frac{1}{2\pi} \int_{-\infty}^{\infty} \!\!dk\, k^{2 n} \ln T(k) \,.
\end{eqnarray}
This set of integral equations relates the BH greybody factors to the KdV integrals. These relations can be recast in the following convenient form      
\begin{eqnarray}
\mu^{}_{2n} = \int_{-\infty}^{\infty} \!\! dk\, k^{2 n}  p(k) \,,
\label{moments-hamburger}
\end{eqnarray}
where
\begin{eqnarray}
\mu^{}_{2n} = (-1)^{n}\frac{I_{2n+1}}{2^{2n+1}} \,.
\label{moments-hamburger-kdv}
\end{eqnarray}
and 
\begin{equation}
p(k) = - \frac{\ln T(k)}{2\pi} \,.
\label{probability-distribution}
\end{equation}
We have left a minus sign in this equation in order to express everything in terms of positive quantities.  With this we have arrived at the main result of this work: The determination of the BH scattering transmission probability $T(k)$ in terms of the BH potential barrier KdV integrals:

~

\noindent{\em BH Moment Problem}. The transmission probability/greybody factors associated with BH scattering processes are uniquely determined in terms of the BH potential KdV integrals via a (Hamburger) moment problem.

~

The key equation is Eq.~\eqref{moments-hamburger}, which clearly defines a \textit{moment problem}~\cite{Shohat1943ThePO,akhiezer1965classical}, i.e. the problem of finding a probability distribution starting from the knowledge of its moment sequence. In this case, the probability distribution is given by $p(x)$ [Eq.~\eqref{probability-distribution}] and the moments by $\mu^{}_{2j}$ [Eq.~\eqref{moments-hamburger-kdv}]. Notice that we only have even moments. This is a key property that we discuss in the next section.  

Let us conclude this section with a remark on Eq.~\eqref{probability-distribution}. We have already mentioned in Sec.~\ref{Sec:kdv-symmetries} that the KdV equation is a completely integrable Hamiltonian system for which one can find action angle variables~\cite{Zakharov:1971faa,Arnold:1989am} $(P(k),Q(k))$. It turns out that the action variable $P(k)$ is associated to the distribution $p(k)$ of the BH moment problem, defined by Eq.~\eqref{probability-distribution}, while the angle variable $Q(k)$ is given by the argument of the Bogoliubov coefficient $b(k)$, that is
\begin{eqnarray}
P(k) = 2k\, p(k) \,,\quad 
Q(k) = {\rm arg}\, b(k) \,.
\end{eqnarray}
This means that the KdV evolution equations for the scattering data~\eqref{ab-kdv-flow} are actually Hamilton's equations for the KdV written in terms of action-angle variables.

\section{Greybody Factors from KdV Integrals: Moment Problem Methods}  \label{Sec:moment-problem}

We have just seen that the greybody factors associated with BH scattering processes are connected to the KdV integrals of the BH potential(s) via a moment problem. Let us now discuss this new point of view to see how one can use it in order to provide practical methods to compute the greybody factors. We illustrate this discussion with a case where all the equations involved can be solved analytically, namely the case of a P\"oschl-Teller potential. 

Let us start by considering a particular interval of the real line,  $\mathcal{I}\subseteq \mathbb{R}$. The moments of a distribution $p(x)$ with support on $\mathcal{I}$ are defined as
\begin{eqnarray}
\label{moment-problem}
\mu^{}_n = \int^{}_{\mathcal{I}} dx\, x^n\, p(x)  \quad
n= 0,1,2,\ldots \,,
\end{eqnarray}
where $x$ here is just a real coordinate on $\mathcal{I}$ and $p(x) dx = d\rho (x)$ defines a positive measure.  We also assume that $\{\mu_n\}$ is a sequence of real numbers. In broad terms, the moment problem can be thought as the inverse problem of finding the probability distribution $p(x)$ from the knowledge of its moments $\{\mu_n\}$.  To that end, we need to consider the three following elements:
\begin{enumerate}
\item Existence: Is there a function $p(x)$ on $\mathcal{I}$ whose moments are given by $\{\mu_n\}$?
\item Uniqueness: Do the moments $\{\mu_n\}$ determine uniquely a distribution $p(x)$ on $\mathcal{I}$?
\item Construction: How can we construct all such probability distributions?
\end{enumerate}
When the solution is unique the problem is said to be {\em determinate}, otherwise it is called {\em indeterminate}. When $\mathcal{I}$ is the positive real line, $\mathcal{I}=[0, \infty)$, the moment problem is known as the {\em Stieltjes} moment problem~\cite{AFST_1894_1_8_4_J1_0,Stieltjes:1993bo}, when  $\mathcal{I}=\mathbb{R}$ it is known as the {\em Hamburger} moment problem~\cite{Hamburger:1919,Hamburger1920, Hamburger1921a, Hamburger1921b} and, finally, when $\mathcal{I}=[a, b]\subset\mathbb{R}$ it is known as a Hausdorff moment problem.  By looking at the definition of our BH moments in Eq.~\eqref{moments-hamburger}, it is clear that the problem we are facing is a special case of the Hamburger one, called the {\em symmetric Hamburger} moment problem, which is characterized by a probability distribution that is symmetric with respect to the origin so that the only non-vanishing moments are the even ones [see Eq.~\eqref{moments-hamburger}]. 

This problem has a long history~\cite{KJELDSEN199319}. It originated with the work of Stieltjes~\cite{AFST_1894_1_8_4_J1_0,Stieltjes:1993bo} on the study of analytic properties of continued fractions, although moments were already investigated by Chebyshev~\cite{Chebyshev:1874} and later by Markov~\cite{Markov:1884} (see~\cite{MR0044591} for an historical account) while studying limiting values of integrals with continued fractions. Nevannlina~\cite{nevanlinna1922asymptotische} made contact with the field of complex functions and Riesz~\cite{Riesz1921,Riesz1922,Riesz1923} with functional analysis. Important contributions to the moment problem were also made by Carleman, Wall and others (see~\cite{Shohat1943ThePO,akhiezer1965classical} for other references). For a more recent account on the connections with different mathematical branches see~\cite{schmudgen2017moment} and references therein. Although many results regarding existence and uniqueness of solutions have been found, no general solution method exists.

For convenience (see the expressions of the KdV integrals shown in Appendix~\ref{App:KdV}), we reformulate the moment problem~\eqref{moments-hamburger} in terms of the dimensionless variable $w = r_s k$:
\begin{eqnarray}
\hat{\mu}^{}_{2n} =
\int_{-\infty}^{\infty}\!\! dw\, w^{2n} p(w) \,,~
\hat{\mu}^{}_{2n} =
r^{2n+1}_s \mu^{}_{2n} \,.
\label{moment-problem-adimensional}
\end{eqnarray}
%

\subsection{Hamburger Moment Problem: Existence and Determinacy Conditions}

\begin{figure}[t]
\centering
\includegraphics[width=0.49\textwidth]{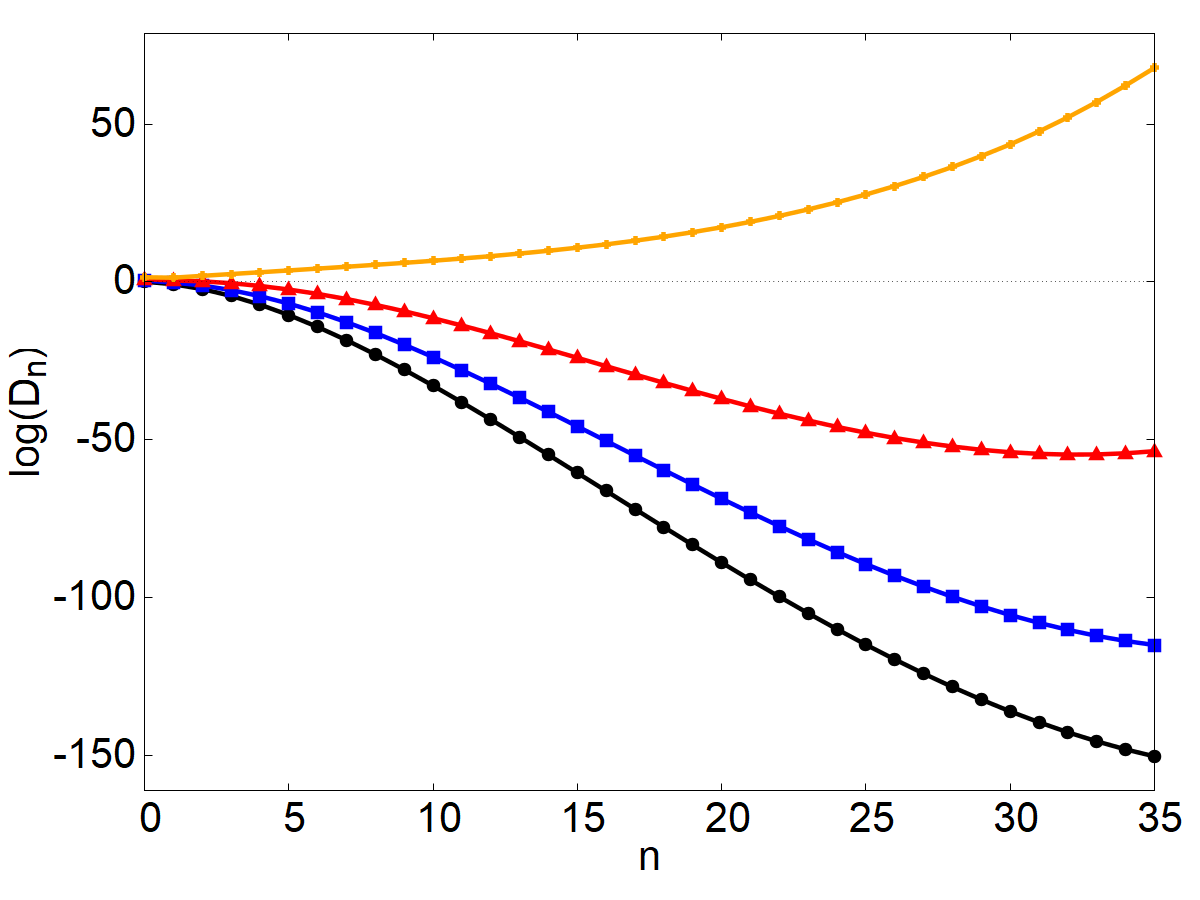}
\caption{Plot of the Hankel determinants up to $n=35$ in logarithmic scale of base $10$, for: $\ell =2$ (black dots), $\ell =3$ (blue squares), $\ell =5$ (red triangles), and $\ell =10$ (orange crosses). The determinants are evaluated using the dimensionless moments of Eq.~\eqref{moment-problem-adimensional}.
\label{hankel}}
\end{figure}

\begin{figure}[t]
\centering
\includegraphics[width=0.49\textwidth]{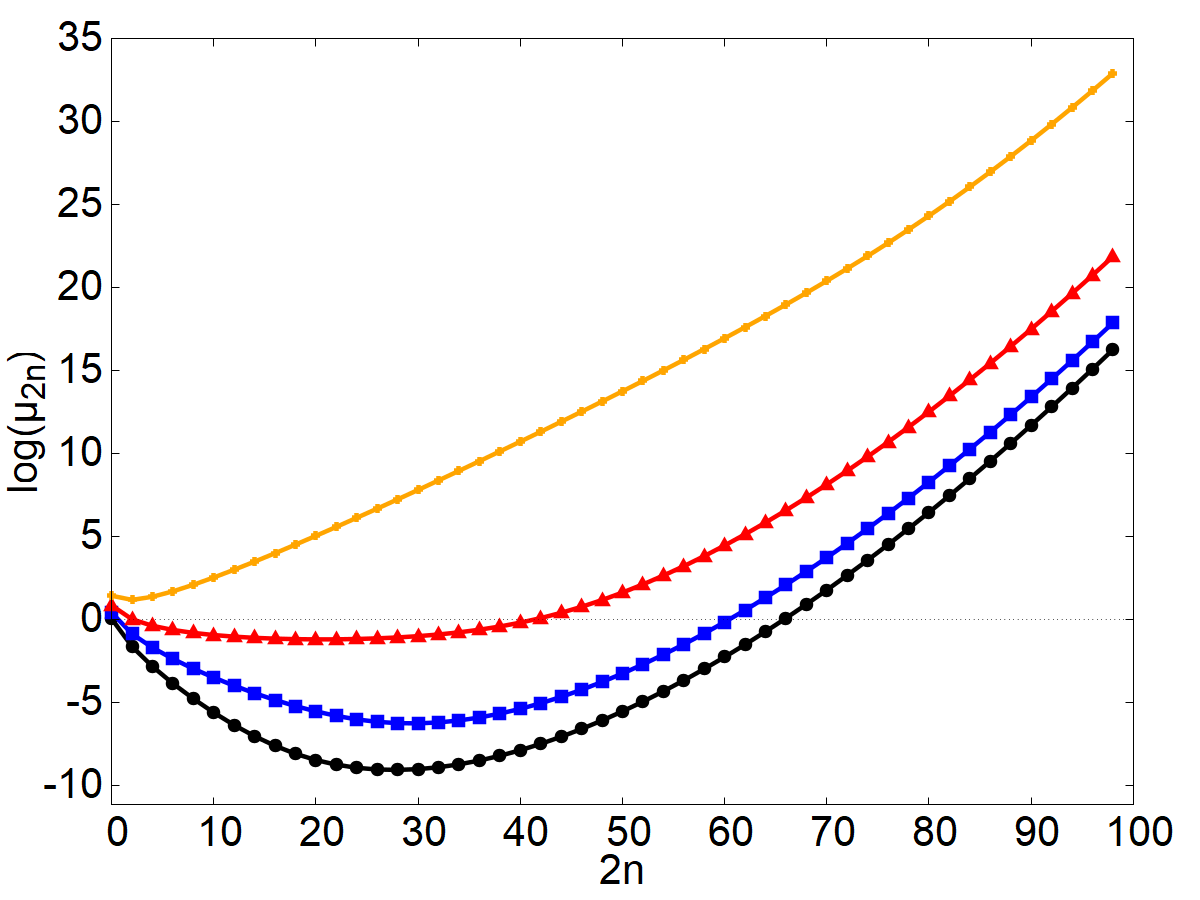}
\caption{Plot, in logarithmic scale of base $10$, of the dimensionless moments [see Eq.~\eqref{moment-problem-adimensional}] for: $\ell =2$ (black dots), $\ell =3$ (blue squares), $\ell =5$ (red triangles), and $\ell =10$ (orange crosses).
\label{moments-RW}}
\end{figure}

Here we study our BH moment problem~\eqref{moments-hamburger} as a Hamburger moment. We start by looking at what can we say about  existence and uniqueness of the solution. We start by introducing the so-called {\em Hankel determinants}, $\{D_n\}$, associated to a real sequence $\{\mu_n\}$ as
\begin{eqnarray}
D^{}_n =
\begin{vmatrix}
\mu^{}_0 & \mu^{}_1 & \mu^{}_2 & \cdots & \mu^{}_n \\
\mu^{}_1 & \mu^{}_2 & \mu^{}_3 & \cdots &\mu^{}_{n+1}\\
\mu^{}_2 & \mu^{}_3 & \mu^{}_4 & \cdots &\mu^{}_{n+2}\\
\vdots & \vdots & \vdots & \cdots & \vdots\\
\mu^{}_{n} & \mu^{}_{n+1} & \mu^{}_{n+2} & \cdots & \mu^{}_{2n}
\end{vmatrix}
\,.
\end{eqnarray}
Existence of solutions to the Hamburger moment problem is guaranteed provided all the Hankel determinants $D_n$ are positive.
For a symmetric Hamburger moment problem like ours, all odd entries  vanish, which makes it easier to check whether the Hankel determinants are positive.
A possible strategy is to consider the Cholesky decomposition (see, e.g.~\cite{Press:1992nr}) to simplify the calculations. In  Fig.~\ref{hankel}, we show the first Hankel determinants for different values of $\ell$. For small $\ell$, the $D_n$ first decrease but then start increasing with $n$, while for bigger $\ell$ the determinants grows rapidly (this appears to be related to the behavior of the moments, see Fig.~\ref{moments-RW}). 
It seems natural to claim that the Hankel determinants will never become negative. Therefore, we conclude that solutions to the problem~\eqref{moments-hamburger} exist.

\begin{figure}[t]
\centering
\includegraphics[width=0.49\textwidth]{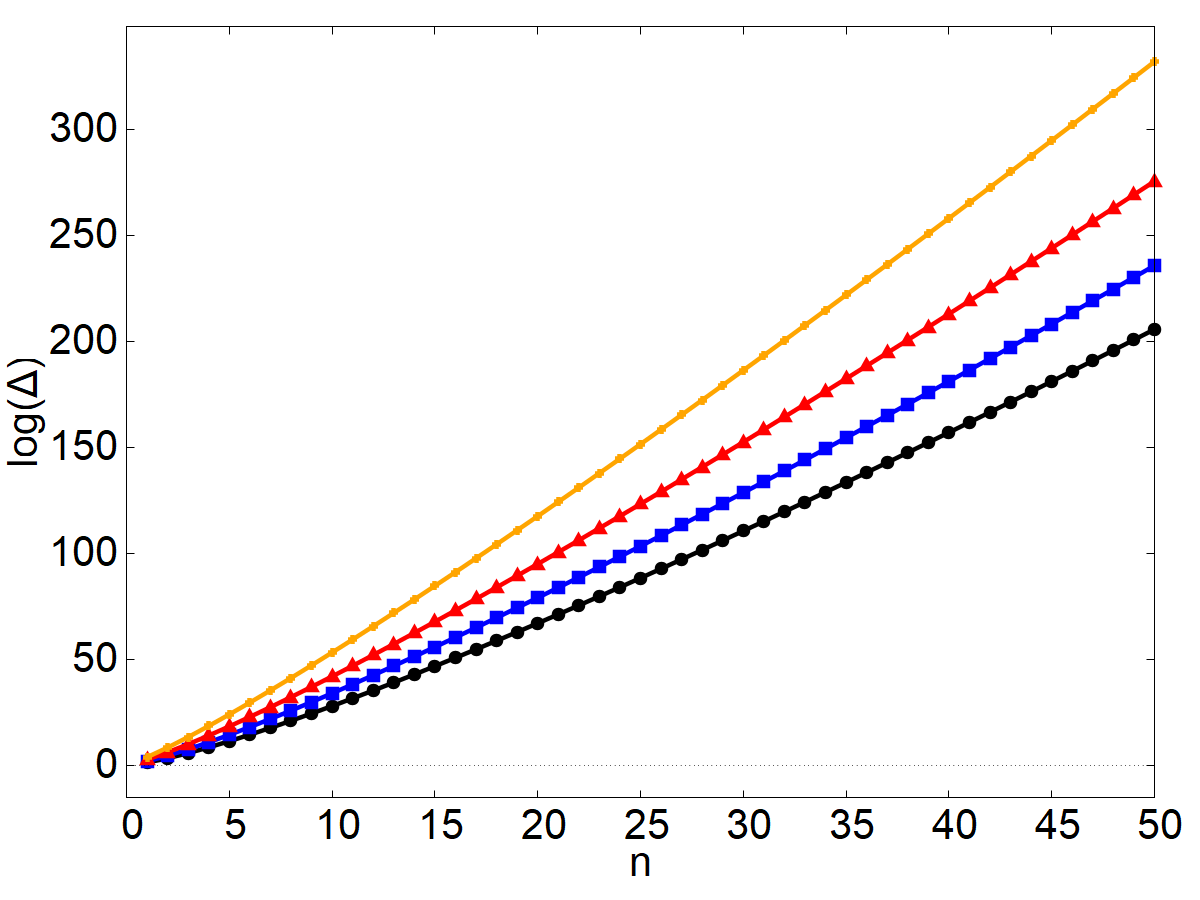}
\caption{Plot of the difference $\Delta(n)$, in logarithmic scale of base $10$, evaluated using the dimensionless moments of Eq.~\eqref{moment-problem-adimensional}. For this plot we use: $\ell =2$ (black dots), $\ell =3$ (blue squares), $\ell =5$ (red triangles), and $\ell =10$ (orange crosses).
\label{uniqueness}}
\end{figure}

The second fundamental condition to check is whether the solution is unique. There are a number of statements defining determinacy~\cite{schmudgen2017moment} of the moment problem. One of the most used ones is the {\em Carleman condition}, which states that a sufficient condition for the Hamburger moment problem to be determinate is
\begin{eqnarray}
\sum_{n=1}^{\infty} \mu_{2n}^{-\frac{1}{2n}} = \infty  \,.
\label{carleman-condition}
\end{eqnarray}
Of course, Carleman's condition has the disadvantage of only providing a sufficient condition, i.e. there can be moment sequences for which this criterion does not hold but they are still determinate. However, it also gives us the following intuitive interpretation: If the even moments do not tend to infinity {\em too quickly}, then the moment problem is determinate. A practical approach to study this growth condition, which is equivalent to Carleman's condition, is to study whether there is a constant $C$ for which
\begin{equation}
\mu^{}_{2n} \leq C^n (2  n)! \equiv \mathcal{C}(n)\,,\;\mbox{for all}\; n > 0 \,.
\label{uniqueness-condition}
\end{equation}
In Appendix~\ref{App:KdV} we show analytic expressions for the KdV integrals for the RW potential (and hence, for any other potential associated with BH perturbations). As we can see, the KdV integrals grow with the harmonic number $\ell$ at most as: $I_{2n+1} \propto L^{n+1}$ where $L=\ell(\ell +1)$ (see also Fig.~\ref{moments-RW}). Furthermore, Eq.~\eqref{moments-hamburger-kdv} shows that $\hat{\mu}_{2n}\propto I_{2n+1}/2^{2n +1}$ so that it seems natural to think that taking $C =(L/2)^2$ will meet the criteria~\eqref{uniqueness-condition}. The factorial term in Eq.~\eqref{uniqueness-condition} already grows faster than any power of $L$ so that it only helps improving the condition. Indeed, notice from Fig.~\ref{uniqueness} that the difference 
\begin{equation}
 \Delta(n)=   \mathcal{C}(n) - \hat{\mu}^{}_{2n}\,,
\end{equation}
is actually increasing, indicating that condition~\eqref{uniqueness-condition} is always satisfied with $C =(L/2)^2$. 
Independently on whether or not there is a more stringent bound, the important result is that the moments~\eqref{moments-hamburger-kdv}, and thus the KdV integrals, uniquely determine the distribution~\eqref{probability-distribution} and therefore the BH greybody factors.

\subsection{Analytic Methods to Solve the Moment Problem}

In order to construct solutions to the moment problem, it is useful to consider the connection with the theory of analytic functions~\cite{nevanlinna1922asymptotische}. Actually, given a solution $p(k)$ of the moment problem, one can define its {\em Stieltjes transform} function~\cite{Shohat1943ThePO,akhiezer1965classical} as 
\begin{eqnarray}
F(z) =
\int_{-\infty}^{\infty}\!\! dk\, \frac{p(k)}{k-z} \,,
\label{stieltjes-transform}
\end{eqnarray}
which is analytic for every $z\in \mathbb{C\setminus R}$. Besides, for any range of angles $\epsilon \leq \arg(z)\leq \pi - \epsilon$ with $0 < \epsilon < \pi/2$, this function is asymptotically represented by the following series
\begin{align}
F(z) = - \sum_{n=0}^{\infty} \frac{\mu_n}{z^{n+1}} \,,
\label{stieltjes-transform-asymptotic}
\end{align}
which results from formally expanding the denominator of Eq.~\eqref{stieltjes-transform} for large $|z|$.
The inverse is also true, i.e. given the function $F(z)$ with those properties, there exists only one distribution $p(k)$ that solves the moment problem with moments $\{\mu_{n}\}$, and such that $F(z)$ has the integral representation~\eqref{stieltjes-transform} with asymptotics given in Eq.~\eqref{stieltjes-transform-asymptotic}. From the series expansion, the Stieltjes transform can also been seen as the generating function for the moments $\mu_n$. Another important consequence of condition~\eqref{uniqueness-condition} is that it guarantees that $F(z)$ actually has the asymptotic representation~\eqref{stieltjes-transform-asymptotic} (see Ref.~\cite{Bender:1978bo}), not just has a formal expansion.

There is a one-to-one relation between the moment problem and the problem of finding the analytic function $F(z)$ in $\mathbb{C\setminus R}$ and with the asymptotic expansion of Eq.~\eqref{stieltjes-transform-asymptotic}. We then conclude that the solution $p(k)$ to the moment problem is determined uniquely by $F(z)$. Indeed, the {\em Stieltjes-Perron inversion formula}~\cite{AFST_1894_1_8_4_J1_0} (see also~\cite{Shohat1943ThePO,akhiezer1965classical,schmudgen2017moment}) provides the inverse relation between $F(z)$ and $p(k)$ and it reads
\begin{eqnarray}
p(k) & = & \lim^{}_{\epsilon\to 0} \frac{F(k+i\epsilon) - F(k-i\epsilon) }{2\pi i} \notag \\
& = &
\lim^{}_{\epsilon\to 0} \frac{\Im F(k+i\epsilon)}{\pi} \,,
\label{stieltjes-perron-inversion}
\end{eqnarray}
where the last equality holds if $\overline{F(z)} = F(\overline{z})$. This feature is guaranteed by the Herglotz property of the Stieltjes transform~\cite{Bender:1978bo}, by which the sign of $\Im{F(z)}$ is the same as the sign of $\Im{z}$. Therefore, the moment problem is equivalent to the problem of inverting the Stieltjes transform~\eqref{stieltjes-transform} with the use of the Stieltjes-Perron inversion formula~\eqref{stieltjes-perron-inversion}, which relates the probability distribution with the branch cut discontinuity of the Stieltjes transform across the real axis.
It is clear that Eq.~\eqref{loga-logr} defines a Stieltjes transform after identifying $p(k)$ with the distribution given by Eq.~\eqref{probability-distribution} and
\begin{equation}
F(k+i\epsilon) = i \ln{a(k+i\epsilon)}
\,.
\label{stieltjes-scattering}
\end{equation}
Therefore, Eq.~\eqref{stieltjes-perron-inversion} now reads 
\begin{eqnarray}
\ln T(k) = - \lim^{}_{\epsilon\rightarrow 0}
2\,\Re \ln{a(k+i\epsilon)} \,.
\label{stieltjes-inversion-loga}
\end{eqnarray}

In order to illustrate how the procedure described above works, we are going to show what happens in a particular case where we can carry out all the steps using analytic techniques, namely the case of a P\"oschl-Teller potential:
\begin{eqnarray}
V(x) = \frac{V^{}_0}{\cosh^2{(\alpha x)}}
= \frac{\alpha^2 \lambda(1-\lambda)}{\cosh^2(\alpha x)} \,.
\label{pt-potential}
\end{eqnarray}
Although this potential does not belong to the family of potentials describing the perturbations of Schwarzschild BHs, it is one of the exactly solvable problems in scattering theory (see Appendix~\ref{App:PT} for details), and as such it is very interesting to consider in order to understand the procedure in an analytical way. 

From Eq.~\eqref{pt-transmission} we can directly obtain the exact expression for $a(k)$ 
\begin{eqnarray}
a(k) =
\frac{\Gamma(1-ik) \Gamma(-ik)}{\Gamma\left(\frac{1}{2}-i(k-l)\right)\Gamma\left( \frac{1}{2}-i(k+l) \right)}
\ ,
\label{ak-pt-barrier}
\end{eqnarray}
Notice that in the left-hand side of Eq.~\eqref{stieltjes-inversion-loga}, $T(k)= |a(k)|^{-2}$, is evaluated on the real line, while $a(k+ i \epsilon)$ on the right-hand side, is the analytical continuation into the complex plane. Therefore, we consider the analytical extension of Eq.~\eqref{ak-pt-barrier} for $k \in \mathbb{C}$, whose existence is guaranteed by the analytic properties of the Gamma function~\cite{lebedev1972special}. The logarithm of $a(k)$ can then be decomposed as
\begin{eqnarray}
&& \ln{a(k)} = \ln \Gamma\left(1-ik\right) + \ln \Gamma\left(-ik\right) 
\nonumber
\\
&-& \ln{\Gamma\left(\frac{1}{2}-i(k-l)\right)} 
-\ln{\Gamma\left(\frac{1}{2}-i(k+l)\right)}
\,.
\end{eqnarray}
After taking the real part of this and the limit for the imaginary part of $k$ going to zero, we end up with the following expression for the transmission probability
\begin{eqnarray}
\ln T(k) =
\ln{\left[\frac{\sinh^2(\pi k)}{\cosh^2(\pi k)+\sinh^2(\pi l)} \right]}
\,,
\end{eqnarray}
where we have used some mathematical properties of the Gamma function~\cite{Abramowitz:1970as}.
Therefore, we have recovered Eq.~\eqref{pt-transmission-real-k} and have proved, in this particular case, that the Stieltjes-Perron inversion formula provides the correct solution to the moment problem. However, given the necessity of the explicit analytical expression of the Stieltjes transform, this approach is not going to be useful in general  cases.

A different approach may be attempted. By looking at Eqs.~\eqref{probability-distribution} and~\eqref{stieltjes-scattering}, it is clear that the asymptotic expansion~\eqref{stieltjes-transform-asymptotic} corresponds to Eq.~\eqref{ak-expansion}, up to an imaginary unit. At this point, one may be tempted to use such expansion in terms of the moments together with the inversion formula~\eqref{stieltjes-perron-inversion} to obtain the greybody factors in terms of the KdV integrals.
However, the Stieltjes-Perron inversion formula depends heavily on the analytic properties of the Stieltjes function as it essentially relates the distribution, on the real line, to the jump of the Stieltjes transform across the real axis (which by definition is discontinuous there). Unfortunately, this behavior seems to be lost when we consider the asymptotic expansion~\eqref{stieltjes-transform-asymptotic}. What is more, Eq.~\eqref{stieltjes-transform-asymptotic} is only valid in the complex plane, so that no such expansion is expected to hold on the real line. We can use again the case of the P\"oschl-Teller potential to show this. The greybody factor $T(k)$ is given in Eq.~\eqref{pt-transmission-real-k}. Imagine to start from the asymptotic expansion of Eq.~\eqref{ak-expansion}. The Stieltjes-Perron inversion formula~\eqref{stieltjes-perron-inversion} will clearly preserve the asymptotic form. However, we immediately see from Eq.~\eqref{pt-transmission-real-k} that the distribution~\eqref{probability-distribution} on the real line does not allow any asymptotic expansion around $k \to \infty$. Moreover, if we just naively substitute the general expansion~\eqref{stieltjes-transform-asymptotic} into Eq.~\eqref{stieltjes-perron-inversion}, everything vanishes in the limit $\epsilon\to 0$. 

Finally, it is important to mention that Padé approximants to the Hamburger series~\eqref{stieltjes-transform-asymptotic} are convergent~\cite{baker_graves-morris_1996}. However, as for the series itself, we cannot take the limit of these Padé approximants to the real line because otherwise Eq.~\eqref{stieltjes-perron-inversion} would give a vanishing result.

\section{Conclusions and Discussion}\label{Conclusions-and-Discussion}

In this work, we have shown that the BH greybody factors are uniquely determined by the set of KdV integrals associated with the BH potential barrier that describes the response of the BH to gravitational perturbations. This result is a consequence of the rich structure and properties of the space of master functions and equations that was presented in previous works~\cite{Lenzi:2021wpc,Lenzi:2021njy}. In these works, it was shown that the space of master functions and equations admits different types of isospectral transformations/symmetries. First of all, DTs connect all possible master functions and equations and preserve the spectrum of QNMs as well as the transmission and reflection coefficients, characterizing the continuous part of the spectrum of the associated time-independent Schr\"odinger operator [See Eq.~\eqref{schrodinger}]. For these reasons, we called Darboux covariance the existence of this symmetry.  On the other hand, the time-independent master equations admit KdV deformations that are isospectral and that leave invariant the Bogoliubov coefficient $a(k)$ (the inverse of the BH transmission coefficient). This conservation law gives rise to the infinite sequence of KdV integrals which, in turn, we have shown to determine completely the BH greybody factors. Moreover, the KdV integrals do not depend on the specific choice of master function/equation, they are the same for all the elements of the space of master functions/equations.

This deep connection between greybody factors, DTs, and KdV integrals has been shown to take place via a moment problem, i.e. the problem of finding a probability distribution given its moments. In our case, the moments are essentially given by the KdV integrals [see Eq.~\eqref{moments-hamburger-kdv}].  Among the different possible moment problems, we have shown that the BH moment problem is of the (symmetric) Hamburger type. 
We have argued that, due to the particular characteristics of the moment sequence, the moment problem admits a solution and that such solution is actually uniquely determined.  However, there is no universal recipe to invert the moment problem and find the distribution given the sequence of moments. Only in a few particular cases, this inversion has been carried out analytically. In particular, when the sequence of moments corresponds to the Catalan numbers~\cite{bostan:hal-02425917}, and also in the case of the P\"oschl-Teller potential~\ref{Sec:moment-problem}, as we have seen in Sec.~\ref{Sec:moment-problem}. 

This work brings a completely new perspective to the study of BH scattering via the moment problem and the mathematical structures that have made it possible to arrive to this result. 
In a subsequent paper~\cite{Lenzi:2022pth}, we present some new approximate semi-analytical techniques for the solution of the BH moment problem (i.e. obtaining the BH greybody factors from the KdV integrals) and compare the results with previous works in the literature, showing that our new methods appear to be quite competitive and have a wide range of applications.

The results of this paper are restricted to the case of Schwarzschild BH scattering processes. But it is important to stress that the conclusions we have obtained are actually quite general. The only requirement is that the problem studied is described by a time-independent Schr\"odinger equation with an integrable potential barrier without bound states. In this sense, the developments of this paper can be extended directly to other physical problems of interest, in particular to rotating Kerr BHs by using the results that come from the Teukolsky master equations and their separable structure~\cite{Teukolsky:1972le,Teukolsky:1972my,Taylor:1972pty}. We can also go further in the development of these techniques to try to come out with a new perspective for the computation of quasinormal modes, which is of great interest in the context of gravitational wave physics. Finally, we can try to transfer these developments to other theories of gravity, with different action principle, and/or different dimensions, and/or extra fields.

%
\begin{acknowledgments}
We thank Jos\'e Lu\'{\i}s Jaramillo for enlightening discussions regarding the dynamics of perturbed black holes.
ML and CFS  are supported by contract PID2019-106515GB-I00/AEI/10.13039/501100011033 (Spanish Ministry of Science and Innovation) and 2017-SGR-1469 (AGAUR, Generalitat de Catalunya). 
This work was also partially supported by the program Unidad de Excelencia Mar\'{\i}a de Maeztu CEX2020-001058-M (Spanish Ministry of Science and Innovation).
\end{acknowledgments}

\appendix

\section{Gauge-Invariant Metric Perturbations}\label{gauge-invariant-mp-harmonics}
Under gauge transformations, spacetime perturbations transform according to Eq.~\eqref{gauge-transformation-metric}. To understand how the different harmonic transform, we have to decompose the generator of the gauge transformation, $\xi^\mu$ [see also Eq.~\eqref{gauge-transformation}], in harmonics (we drop out the harmonic indices $(\ell,m)$):
\begin{equation}
\xi^{}_{\mu} dx^{\mu} = \left\{ \begin{array}{ll}
r^{2}\gamma X^{}_{A}d\Theta^{A} \,, & \mbox{odd parity}\\[2mm]
\alpha^{}_{a}dx^{a} + r^{2}\beta Y^{}_{A}d\Theta^{A} \,, & \mbox{even parity}
\end{array} \right.
\end{equation}
where the gauge functions $(\gamma,\alpha_a,\beta)$ are functions of the $M_2$ coordinates $\{x^a\}$. Then, odd-parity metric perturbations transform as
\begin{eqnarray}
h'^{}_{a} & =  & h^{}_{a} - r^{2}\gamma^{}_{:a} \,, \\
h'^{}_{2} & =  & h^{}_{2} - 2\, r^{2}\gamma \,.
\end{eqnarray}
and even-parity metric perturbations transform as
\begin{eqnarray}
h'^{}_{ab} & =  & h^{}_{ab} - 2\,\alpha^{}_{(a:b)} \,, \\
\jeven'^{}_a & = & \jeven^{}_a - \left(\alpha^{}_{a} + r^{2}\beta^{}_{:a} \right)\,, \\
K' & = & K + \ell(\ell+1)\beta^{} -2\frac{r^{:a}}{r}\alpha^{}_a \,, \\
G' & = & G - 2\beta\,.
\end{eqnarray}
From this transformations we find that, for the odd-parity sector, we have two gauge-invariant quantities:
\begin{equation}
\tilde{h}^{}_{a} = h^{}_{a} -\frac{1}{2}h^{}_{2:a} + \frac{r^{}_{a}}{r} h^{}_{2} \,,
\end{equation}
and there four more for even-parity perturbations:
\begin{eqnarray}
\tilde{h}^{}_{ab} & = & h^{}_{ab} - \kappa^{}_{a:b} - \kappa^{}_{b:a}\,, 
\label{expression-hathab} \\
\tilde{K}         & = & K + \frac{\ell(\ell+1)}{2} G - 2\frac{r^{a}}{r}\kappa^{}_{a} \,,
\label{expression-hatK}
\end{eqnarray}
where
\begin{eqnarray}
\kappa^{}_{a} & = & \jeven^{}_{a} - \frac{r^{2}}{2}G^{}_{:a}\,, \quad
r^{}_a = r^{}_{:a}~\Rightarrow~r^a=g^{ab}r^{}_b\,.
\end{eqnarray}
%

\section{Hamiltonian Formalism in the Infinite–Dimensional Case}
\label{Hamiltonian-Formalism}

In this case we have a continuous of dynamical variables, here labeled by the coordinate $x$.
The changes with respect to the finite-dimensional case are that trajectories are replaced by smooth functions $V(\tau,x)$ and phase space functions are replaced by functionals. A functional of $V$ is defined via integration as follows:
\begin{equation}
F[V] = \int^{\infty}_{-\infty} dx\,f(V,V^{}_{,x}, V^{}_{,xx}, \ldots ) \,. 
\end{equation}
Functional (Fr\`echet) derivatives are defined as~\cite{Olver1998:pj}:
\begin{eqnarray}
\frac{\delta F}{\delta V(x)} & = & \sum_{n=0}^\infty (-1)^n \frac{\partial^n}{\partial x^n}
\frac{\partial f}{\partial V^{(n)}} \nonumber \\
& = & \frac{\partial f}{\partial V} - \frac{\partial}{\partial x}\frac{\partial f}{\partial V^{}_{,x}} + \frac{\partial^2}{\partial x^2}\frac{\partial f}{\partial V^{}_{,xx}} + \ldots  \,,
\end{eqnarray}
where $V^{(n)} = d^nV(x)/dx^n$ and one has to take into account that
\begin{equation}
\frac{\delta V(x)}{\delta V(x')} = \delta(x-x') \,,
\end{equation}
where $\delta(x)$ denotes the Dirac delta distribution.  We can build  Poisson brackets as
\begin{equation}
\left\{ F , G \right\} = \int^{\infty}_{-\infty}\!\!\!\! dx\! \int^{\infty}_{-\infty} \!\!\!\! dx'  \omega(x,x',V) \frac{\delta F}{\delta V(x)} \frac{\delta G}{\delta V(x')} \,,
\end{equation}
where $\omega(x,x',V)$, the symplectic form, has to be such that the Poisson bracket is antisymmetric.  A common choice is~\cite{gardner1971korteweg} 
\begin{equation}
\omega(x,x',V) = \frac{1}{2}\left( \partial^{}_x - \partial^{}_{x'} \right) \delta(x-x') \,.
\label{GZF-form}
\end{equation}
Since the partial derivative operator $\partial_x$ is anti-self-adjoin with respect to the inner scalar product
\begin{equation}
\left( V, W \right) = \int^{\infty}_{-\infty} dx\,V(x) W(x)\,,
\end{equation}
the Poisson bracket can be written as:
\begin{equation}
\left\{ F , G \right\}^{}_{\rm GZF} =   \int^{\infty}_{-\infty} dx\, \frac{\delta F}{\delta V(x)} {\cal D}\frac{\delta G}{\delta V(x)} \,,
\label{GZF-bracket}
\end{equation}
where
\begin{equation}
{\cal D} = \frac{\partial}{\partial x} \,.    
\end{equation}
the Poisson bracket in Eq.~\eqref{GZF-bracket} is sometimes referred to as the Gardner-Zakharov-Faddeev bracket~\cite{gardner1971korteweg,Zakharov:1971faa}.
In this way, Hamilton's equations can be written as follows:
\begin{equation}
\frac{\partial V}{\partial\tau}= \left\{ V , {\cal H}[V] \right\}^{}_{\rm GZF}  = {\cal D} \frac{\delta {\cal H}[V]}{\delta V(x)} \,.
\end{equation}
A different choice of symplectic form was introduced by Magri~\cite{Magri:1977gn}
\begin{eqnarray}
\omega(x,x',V) & =& \left[ - \frac{1}{2}\left(\partial^{3}_x - \partial^{3}_{x'}\right) \right. \nonumber \\
& + & \left. 2\left(V(x)\partial^{}_x - V(x')\partial^{}_{x'}\right) \right] \delta(x-x') \,.
\label{Magri-form}
\end{eqnarray}
This leads to the following Poisson bracket:
\begin{equation}
\left\{ F , G \right\}^{}_{\rm M} = \int^{\infty}_{-\infty} dx\, \frac{\delta F}{\delta V(x)}{\cal E}^{}_V \frac{\delta G}{\delta V(x)} \,,
\end{equation}
where ${\cal E}^{}_V$ is the following operator
\begin{equation}
{\cal E}^{}_V =  -\frac{\partial^3}{\partial x^3} +4V(x)\frac{\partial}{\partial x} + 2 V^{}_{,x}(x) \,,
\end{equation}
It is known that the Magri bracket is closely related to the Virasoro algebra~\cite{GERVAIS1985279}.

\section{1D Scattering from a Potential Barrier}
\label{App:1d-scattering}

We provide some basic elements of one-dimensional scattering theory~\cite{Faddeev:1959yc, Faddeev:1976xar, Deift:1979dt, Novikov:1984id, Ablowitz:1981jq}, assuming an integrable potential barrier without bound states, only with continuous spectrum. We start by considering a  set of independent solutions to the time-independent Schr\"odinger equation~\eqref{schrodinger} near $x \rightarrow \infty$ (spatial infinity for Schwarzschild BHs)
\begin{eqnarray}
f^{}_1 (x,k) = e^{i k x} + {\cal O}(1) \,,\;
f^{}_2 (x,k) = e^{-i k x} + {\cal O}(1) \,,
\label{boundary-right}
\end{eqnarray}
and a second set satisfying analogous conditions at $x \rightarrow -\infty$ (the BH horizon)
\begin{eqnarray}
g^{}_1 (x,k) = e^{i k x} + {\cal O}(1) \,,\;
g^{}_2 (x,k) = e^{-i k x} + {\cal O}(1) \,.
\label{boundary-left}
\end{eqnarray}
Since the potential is real we have
\begin{eqnarray}
f^{}_1 (x,k) = \overline{f}^{}_{2}(x,k)\,, \quad
g^{}_1 (x,k) = \overline{g}^{}_{2}(x,k) \,,
\label{ffbar}
\end{eqnarray}
and, in addition, 
\begin{eqnarray}
f^{}_1(x,k) = f^{}_2(x,-k) \,, \quad
g^{}_1(x,k) = g^{}_2(x,-k) \,.
\label{ff-k}
\end{eqnarray}
Therefore, the pairs $\left\{f(x, k)\,,\, \overline{f}(x, k)\right\}$ and $\left\{g(x, k)\,,\,  \overline{g}(x, k)\right\}$, where $f_1\equiv f$ and $g_1 \equiv g$, form two fundamental sets of solutions (for $k \neq 0$), known as the \textit{Jost solutions}~\cite{Taylor:1972pty,Newton:1982qc}. 

The time-independent Schr\"odinger equation, with boundary conditions given by Eqs.~\eqref{boundary-right} and~\eqref{boundary-left}, is equivalent to the following Volterra-type integral equations (see~\cite{Faddeev:1976xar,Deift:1979dt})
\begin{eqnarray}
f(x, k) & = & e^{i k x} + \int_{-\infty}^{\infty} \!\!\! dy\,
G^{}_1(x-y,k) V(y) f(y,k)\,,
\label{volterra-f} \\
\overline{g} (x, k) & = & e^{-i k x} + \int_{-\infty}^{\infty}\!\!\! dy\, G^{}_2(x-y,k) V(y) \overline{g}(y, k) \,,
\label{volterra-g}
\end{eqnarray}
where $G_1$ and $G_2$ are Green's functions given by
\begin{eqnarray}
G^{}_1(x-y,k) & = & -\theta(y-x) \frac{\sin{\left[k(x-y) \right]}}{k}\,,      \label{green-G1} \\ 
G^{}_2(x-y,k) & = & \theta(x-y) \frac{\sin{\left[k(x-y) \right]}}{k} \,, \label{green-G2}
\end{eqnarray}
and $\theta(x)$ is the Heaviside theta function, defined as
\begin{eqnarray}
\theta(x) = \left\{ \begin{array}{lc}
1  &  \mbox{for~}x>0\,, \\[1mm]
0  &  \mbox{for~}x<0\,.
\end{array} \right.
\end{eqnarray}
Starting from the Volterra equations one can: prove that the Jost functions are solutions of~\eqref{schrodinger} using the method of successive approximations~\cite{Faddeev:1976xar,Deift:1979dt}; show the linear independence of the pairs of solutions $\left\{ f(x,k), \overline{f}(x,k) \right\}$, $\left\{g(x,k), \overline{g}(x,k) \right\}$; and study their analytic properties in the complex $k$-plane.
Indeed, the Wronskians read 
\begin{eqnarray}
W\left[f\,, \overline{f}\right] =
W\left[g\,, \overline{g}\right] = 2 i k \,,
\label{wronskians}
\end{eqnarray}
where we just considered Eqs.~\eqref{volterra-f} and~\eqref{volterra-g} and the Wronskian~\eqref{wronskian-def}.
They are different form zero whenever $k\neq 0$, thus we have two pairs of linearly independent solutions which we can express as linear combinations of each other, that is,
\begin{eqnarray}
f(x, k) & = & a(k) g(x,k) + b(k) \overline{g}(x,k)
\,, \label{asymptotic-basis-f} \\[1mm]
\overline{f}(x,k) & = & \overline{b}(k) g(x,k) + \overline{a}(k) \overline{g}(x,k) \,.
\label{asymptotic-basis-fbar}
\end{eqnarray}
Here, we identify the \textit{transition matrix} $\mathcal{T}$
\begin{eqnarray}
\mathcal{T} = \begin{pmatrix}
a(k) & b(k) \\
\overline{b}(k) & \overline{a}(k) 
\end{pmatrix} \,,
\end{eqnarray}
which relates solutions with asymptotics at $x\rightarrow-\infty$ to solutions with asymptotics at $x\rightarrow+\infty$. The coefficients $a(k),b(k),\overline{a}(k),\overline{b}(k)$ can be expressed in terms of the Wronskians of the solutions
\begin{eqnarray}
a(k) & = & \frac{W[f\,,\overline{g}]}{2 i k} \,,\quad
b(k) = \frac{W[g\,,f]}{2 i k}\,,  \label{wronskians-ab} \\
\overline{a}(k) & = &  \frac{W[g\,,\overline{f}]}{2 i k}\,,\quad
\overline{b}(k)  = \frac{W[\overline{f}\,, \overline{g}]}{2 i k} \,,
\label{wronskians-ab-bar}
\end{eqnarray}
so that, using Eqs.~\eqref{ffbar} and~\eqref{ff-k}, we have 
\begin{eqnarray}
a(-k) = \overline{a}(k)\,,\quad  b(-k) = \overline{b}(k) \,.
\end{eqnarray}
After replacing Eqs.~\eqref{asymptotic-basis-f} and~\eqref{asymptotic-basis-fbar} in the first equality of Eq.~\eqref{wronskians} we obtain
\begin{eqnarray}
|a(k)|^2 - |b(k)|^2 = 1 \,.
\label{unimod-T}
\end{eqnarray}
The physical properties of scattering processes are described by the \textit{scattering matrix}, $\mathcal{S}$, which relates outgoing and ingoing states with respect to the barrier.  It is completely determined by the components of $\mathcal{T}$. Let us consider the outgoing states $f$ and $\overline{g}$ as functions of the the ingoing $\overline{f}$ and $g$ states:
\begin{eqnarray}
f(x,k) & = & \frac{1}{\overline{a}(k)}g(x,k) + \frac{b(k)}{\overline{a}(k)} \overline{f}(x,k) \\
\overline{g}(x,k) & = & - \frac{\overline{b}(k)}{\overline{a}(k)}g(x,k) + \frac{1}{\overline{a}(k)} \overline{f}(x,k) \,,
\end{eqnarray}
where we used Eqs.~\eqref{asymptotic-basis-f} and~\eqref{asymptotic-basis-fbar}. The scattering matrix $\mathcal{S}$ reads 
\begin{eqnarray}
\mathcal{S} = \begin{pmatrix}
\frac{1}{\overline{a}(k)} & \frac{b(k)}{\overline{a}(k)} \\[2mm]
-\frac{\overline{b}(k)}{\overline{a}(k)} & \frac{1}{\overline{a}(k)}
\end{pmatrix} \,.
\end{eqnarray}
Consider now the solution 
\begin{eqnarray}
f(x,k) = \left\{ \begin{array}{lc}
a(k) e^{i k x} + b(k)e^{-i k x}  &  \mbox{for~}x\to -\infty\,, \\[3mm]
e^{i k x}  &  \mbox{for~}x \to \infty\,, 
\end{array} \right.
\label{plane-wave-ab}
\end{eqnarray}
which corresponds to the situation in Figure~\ref{BH-scattering-potential}, where $f(x,k)/a(k)$ represents an incoming plane wave from $x\rightarrow-\infty$. The coefficients of the transition matrix define the transmission and reflection coefficients
\begin{eqnarray}
t(k) = \frac{1}{a(k)} \,, \quad 
r(k) = \frac{b(k)}{a(k)} \,,
\label{reflection-transmission-coefficient}
\end{eqnarray}
and then
\begin{eqnarray}
\frac{f(x,k)}{a(k)} = \left\{ \begin{array}{lc}
e^{i k x} + r(k) e^{-i k x}  &  \mbox{for~}x\to -\infty\,, \\[3mm]
t(k)e^{i k x}  &  \mbox{for~}x \to \infty\,, 
\end{array} \right.
\label{plane-wave-rt}
\end{eqnarray}
The transmission and reflection probabilities, also known as greybody factors, are given by the modulus square of the corresponding coefficients for real $k$
\begin{equation}
T(k)=|t(k)|^2\,\quad R(k)=|r(k)|^2  \,.
\label{rt-probabilities}
\end{equation}
Looking at Eq.~\eqref{unimod-T} we notice that the scattering matrix is unitary, i.e. $\mathcal{S}^{\dagger}\mathcal{S} = \mathcal{S} \mathcal{S}^{\dagger}=1$, and 
\begin{eqnarray}
T(k) + R(k) = 1 \,.
\label{T+R}
\end{eqnarray}
Let us now rewrite Eq.~\eqref{volterra-f} in terms of the function $\chi = f e^{-i k x}$~\cite{Novikov:1984id} as follows
\begin{eqnarray}
\chi(x, k) = 1 + \int_{x}^{\infty}\!\!dy\,
\frac{e^{- 2i k (x-y)} - 1}{2 i k}\, V(y)\,\chi(y,k) \,.
\label{hola}
\end{eqnarray}
where we have used Eq.~\eqref{green-G1}. Since we assume the potential to be integrable, $\chi(x, k)$ admits an analytical continuation into $\Im(k) > 0$, and so does the Jost function $f = \chi e^{i k x}$. The same procedure leads to the conclusion that $\overline{g}$ can also be analytically continued into $\Im(k) > 0$ while $\overline{f}$ and $g$ can be continued into the lower half-plane, i.e. $\Im(k) < 0$. 

Finally, Eqs.~\eqref{wronskians-ab} and~\eqref{wronskians-ab-bar} show that: (i) $a(k)$ can be completely expressed in terms of $f$ and $\overline{g}$, and hence it can be analytically continued into the region $\Im(k) > 0$, having the following asymptotic behavior
\begin{eqnarray}
a(k) = 1 + {\cal O}(1/k) \; \mbox{for}\; |k|\rightarrow\infty\,, 
\; \Im(k) > 0 \,.
\label{asymptotics-a}
\end{eqnarray}
(ii) $\overline{a}(k)$ can be expressed in terms of $\overline{f}$ and $g$ and therefore, it can be analytically continued into the region $\Im(k) < 0$. (iii) The integrability of the potential is not enough to guarantee the analyticity of $b$ and $\overline{b}$. 
%
%
\section{The P\"oschl-Teller Potential} \label{App:PT}
The time-independent Schr\"odinger equation~\eqref{schrodinger} for the case of a P\"oschl-Teller potential [see Eq.~\eqref{pt-potential}] can be solved~\cite{landau1981quantum,Cevik:2016mnr} by using a new coordinate, $y = \tanh(\alpha x)$, and use a new unknown
\begin{eqnarray}
\psi(y) = 2^{ik/\alpha} \left( 1-y^2 \right)^{-ik/2\alpha} \nu(y) \,,
\label{nu}
\end{eqnarray}
such that it satisfies the Jacobi equation
\begin{eqnarray}
&&\left( 1 - y^2 \right) \nu''(y)  +
2 y \left(\frac{i k }{\alpha} -1 \right) \nu'(y) \nonumber \\
&&- \left[ \lambda(1-\lambda) -\frac{k^2}{\alpha^2} -\frac{i k}{\alpha} \right] \nu(y) = 0 \,.
\end{eqnarray}
Introducing another coordinate, $z = (1-y)/2$, the Jacobi equation becomes a hypergeometric equation~\cite{lebedev1972special, Abramowitz:1970as}
\begin{eqnarray}
\nonumber
&z& (1-z) \nu''(z) 
+
\left[\left(1-\frac{i k}{\alpha} \right) - 2\left(1-\frac{i k}{\alpha}  \right)z \right]  \nu'(z) \\
&-&
\left(\lambda -\frac{i k}{\alpha} \right) \left(1- \lambda -\frac{i k}{\alpha}   \right) \nu(z) = 0 \,.
\end{eqnarray}
The general solution reads
\begin{widetext}
\begin{eqnarray}
\nonumber
\psi(x,k)
&=&
A\, 2^{ik/\alpha} \left[1 - \tanh^2(\alpha x)\right]^{-ik/2\alpha}
{}_{2}F_1 \left(\lambda -\frac{i k}{\alpha}, 1- \lambda -\frac{i k}{\alpha}  , 1-\frac{i k}{\alpha};\frac{1 - \tanh(\alpha x)}{2} \right)
\\
&+&
B\, \left[1 - \tanh(\alpha x)\right]^{ik/2\alpha} 
\left[1 + \tanh(\alpha x)\right]^{-ik/2\alpha}
{}_{2}F_1 \left(\lambda, 1- \lambda , 1+\frac{i k}{\alpha};\frac{1 - \tanh(\alpha x)}{2} \right)
\ .
\end{eqnarray}
\end{widetext}
The reflection and transmission coefficients can be found from the asymptotics of the hypergeometric function~\cite{Abramowitz:1970as}. First, the asymptotic behavior for $\psi(x,k)$ is
\begin{eqnarray}
\psi (x,k) = \left\{ \begin{array}{lc}
A\,e^{ikx} + B\, e^{-ikx}  &~x\rightarrow \infty\,, \\[2mm]
A'\,e^{ikx} + B'\, e^{-ikx}  &~ x\rightarrow -\infty\,,
\end{array} \right. 
\label{pt-asymptotic-solution}
\end{eqnarray}
where
\begin{eqnarray}
A' & = & A \,\frac{\Gamma(1-\frac{i k}{\alpha})\Gamma(-\frac{i k}{\alpha})}{\Gamma(\lambda -\frac{i k}{\alpha})\Gamma(1- \lambda -\frac{i k}{\alpha})}
\nonumber
\\ \label{A'}
&+&
B\, \frac{\Gamma(1+\frac{i k}{\alpha})\Gamma(-\frac{i k}{\alpha})}{\Gamma(\lambda )\Gamma(1- \lambda)}
\\
B' & = & A\, \frac{\Gamma(1-\frac{i k}{\alpha})\Gamma(\frac{i k}{\alpha})}{\Gamma(\lambda )\Gamma(1- \lambda)}
\nonumber
\\ \l\label{B'}
&+&
B \,\frac{\Gamma(1+\frac{i k}{\alpha})\Gamma(\frac{i k}{\alpha})}{\Gamma(\lambda +\frac{i k}{\alpha})\Gamma(1- \lambda +\frac{i k}{\alpha})}
\ .
\end{eqnarray}
The physical situation of Fig.~\ref{BH-scattering-potential} requires
\begin{eqnarray}
A=1\,,\quad B=0\,,\quad A'=a(k)\,,\quad B'= b(k)
\,,
\end{eqnarray}
so that
\begin{eqnarray}
a(k) & = & \frac{\Gamma(1-ik) \Gamma(-ik)}{\Gamma(\lambda-ik)\Gamma(1-\lambda-ik)}
\\
b(k) & = & \frac{\Gamma(1-ik) \Gamma(ik)}{\Gamma(\lambda)\Gamma(1-\lambda)}
\ ,
\end{eqnarray}
where we set $\alpha =1$ to simplify the notation. Then, the transmission and reflection coefficients~\eqref{reflection-transmission-coefficient} are
\begin{eqnarray}
t(k) &=& \frac{\Gamma(\lambda-ik)\Gamma(1-\lambda-ik)}{\Gamma(1-ik) \Gamma(-ik)}
\\
r(k) &=&\frac{\Gamma(ik)\Gamma(\lambda-ik)\Gamma(1-\lambda-ik)}{\Gamma(-ik)\Gamma(1-\lambda)\Gamma(\lambda) }
\ .
\end{eqnarray}
When the P\"oschl-Teller potential is a high barrier, i.e. when $\lambda= \frac{1}{2}+i l$ with $l>0$~\cite{Cevik:2016mnr}, the transmission coefficient reads
\begin{equation}
t(k) = \frac{\Gamma\left(\frac{1}{2}-i(k-l)\right)\Gamma\left(\frac{1}{2}-i(k+l)\right)}{\Gamma(1-ik) \Gamma(-ik)}
\,.
\label{pt-transmission}
\end{equation}
Then, the transmission probability, or greybody factor, can be evaluated by taking the square of the modulus of Eq.~\eqref{pt-transmission}:
\begin{eqnarray}
T(k) =
\frac{\sinh^2(\pi k)}{\cosh^2(\pi k)+\sinh^2(\pi l)}
\,,
\label{pt-transmission-real-k}
\end{eqnarray}
where here $k \in \mathbb{R}$. 
%
\section{KdV Integrals for the Regge-Wheeler Potential} \label{App:KdV}
The first KdV integrals for the Regge-Wheeler potential read (where we use the definition: $L= \ell(\ell +1)\,$):
\begin{widetext}
\begin{eqnarray}
I^{}_1 & = & \frac{1}{r^{}_s} \left(L-\frac{3}{2} \right) \,, 
\\
I^{}_3 & = & \frac{-1}{r^{3}_s} \left( \frac{1}{12}L^2 - \frac{3}{10}L + \frac{3}{10} \right) \,, 
\\
I^{}_5 & = & \frac{1}{r^{5}_s} \left( \frac{2}{105} L^3 -\frac{83}{840}L^2 + \frac{5}{28}L - \frac{29}{280} \right)\,,
\\
I^{}_7 & = & \frac{-1}{r^{7}_s} \left( \frac{1}{168}L^4 - \frac{53}{1386}L^3 + \frac{263}{2772}L^2 - \frac{147}{1430}L + \frac{222}{5005} \right)\,,
\\
I^{}_{9} & = & \frac{1}{r^{9}_s}\left( \frac{14}{6435}L^5 - \frac{41}{2574}L^4 + \frac{8}{165}L^3 - \frac{2557}{34320}L^2 + \frac{1203}{19448}L -\frac{723}{38896} \right) \,, 
\\
I^{}_{11} & = & \frac{-1}{r^{11}_s}\left( \frac{1}{1144}L^6 - \frac{67}{9724}L^5 + \frac{73}{2992}L^4 - \frac{186859}{3879876}L^3 + \frac{2253239}{38798760}L^2 - \frac{19129}{587860}L + \frac{64759}{5173168} \right) \,, 
\\
I^{}_{13} & = & \frac{1}{r^{13}_s}\left( \frac{11}{29393}L^7 - \frac{1529}{503880}L^6 + \frac{3091}{251940}L^5 - \frac{103699}{3527160}L^4 + \frac{2719789}{60843510}L^3 - \frac{8348939}{243374040}L^2 + \frac{39}{1292}L \right. \nonumber \\
& + & \left. \frac{4803}{772616} \right) \,, \\
I^{}_{15} & = & \frac{-1}{r^{15}_s}\left( \frac{13}{77520}L^8 -\frac{13}{9690}L^7 + \frac{91}{14535}L^6 - \frac{28639}{1671525}L^5 + \frac{81283}{2674440}L^4 - \frac{7109}{222870}L^3 + \frac{111529}{2228700}L^2 + \frac{17637}{566950}L \right. \nonumber \\
& + & \left. \frac{1194813}{21544100} \right)\,, \\
I^{}_{17} & = & \frac{1}{r^{17}_s}\left( \frac{26}{334305}L^9 - \frac{131}{222870}L^8 + \frac{16}{4845}L^7 - \frac{217519}{23401350}L^6 + \frac{750447}{37702175}L^5 - \frac{1106101}{37702175}L^4 + \frac{83799194}{1676927175}L^3 \right. \nonumber \\
& + & \left. \frac{18613463771}{176316914400}L^2 + \frac{1930951511}{7260108240}L + \frac{83062407817}{246843680160} \right)\,, \\
I^{}_{19} & = & \frac{-1}{r^{19}_s} \left( \frac{17}{458850}L^{10} 
+ \frac{1003}{3991995}L^9 + \frac{48841}{26613300}L^8 - \frac{886822}{206253075}L^7 + \frac{148953983}{9900147600}L^6 - \frac{1144342879}{54450811800}L^5 \right. \nonumber \\ 
& + & \left. \frac{600058031}{24200360800}L^4 + \frac{19712925427}{163352435400}L^3 + \frac{290243073581}{326704870800}L^2 - \frac{43168543419}{21880401400}L + \frac{4524494561511}{1662910506400}
\right) \,.
\end{eqnarray}
\end{widetext}
%
%
%

\end{document}